\documentclass[a4paper, twoside, fontsize=9pt, twocolumn]{scrartcl}
\pdfoutput=1

\usepackage[a4paper, top=88pt, bottom=88pt, left=50pt, right=50pt, headsep=16pt, footskip=28pt]{geometry}
\usepackage{amsfonts,amssymb,amsmath,amsthm,amstext,amssymb,amsopn,mathtools,nicefrac,xfrac}
\usepackage[authoryear,sort,round]{natbib} % for bibliography
\usepackage[utf8]{inputenc}                 % allow utf-8 input
\usepackage{booktabs}                       % professional-quality tables
\usepackage{nicefrac}                       % compact symbols for 1/2, etc.
\usepackage{microtype}                      % microtypography (fixes wording in paragraphs)
\usepackage{graphics,graphicx}              % standard graphics
\usepackage{subcaption}                     % subcaption in figures
\usepackage{enumitem}
\usepackage{lipsum}  

\usepackage[boxruled,linesnumbered]{algorithm2e} % specific algorithm format
\SetKwComment{Comment}{\%}{}
\SetKwInput{KwInput}{Input}
\SetKwInput{KwOutput}{Output}

\usepackage{verbatim}                       % write code directly
\usepackage{appendix}                       % builds an appendix if needed
\usepackage{bm}                             % math fonts
\usepackage{array}                          % write vectors and matrices
\newcolumntype{C}[1]{>{\centering\arraybackslash}m{#1}}  %for centering table text
\usepackage[usenames,dvipsnames]{xcolor}    % more colours (for hyperlinks)
\usepackage{hyperref}                       % hyperlinks
\hypersetup{colorlinks=true,linkcolor=Maroon,filecolor=Magenta,urlcolor=Blue,citecolor=RoyalBlue}
\usepackage[noabbrev,capitalise,nosort,nameinlink]{cleveref}  % uses cref instead of ref
\usepackage{bbm}                            % special math symbols 
\usepackage{mathtools}                      % special maths commands

\newcommand*{\arXiv}[1]{\bgroup\color{blue}\href{https://arxiv.org/abs/#1}{arXiv:#1}\egroup}
\newcommand*{\doi}[1]{\bgroup\color{blue}\href{https://doi.org/#1}{doi:#1}\egroup}
\newcommand*{\email}[1]{\bgroup\color{blue}\href{mailto:#1}{#1}\egroup}
\renewcommand*{\url}[1]{\bgroup\color{blue}\href{#1}{#1}\egroup}
\usepackage{enumitem, moreenum}
\setlist[enumerate]{nosep}
\setlist[itemize]{nosep}
\usepackage{mleftright} \mleftright
\renewcommand{\qedsymbol}{$\blacksquare$}
\renewenvironment{proof}[1][\proofname]{\noindent{\bfseries\sffamily #1.} }{\hfill\qedsymbol\medskip}
\usepackage[textfont={small}, labelfont={sf,bf,small},format=plain,indention=0cm]{caption}
\DeclareCaptionLabelSeparator{figlabelsep}{\,\,\,}
\usepackage{scrlayer-scrpage, xhfill}
\automark[section]{section}
\setkomafont{pageheadfoot}{\normalcolor\sffamily}
\setkomafont{pagenumber}{\normalfont\normalsize\sffamily}
\clearpairofpagestyles
\let\oldtitle\title
\renewcommand{\title}[1]{\oldtitle{#1}\newcommand{\theshorttitle}{#1}}
\newcommand{\shorttitle}[1]{\renewcommand{\theshorttitle}{#1}}
\let\oldauthor\author
\renewcommand{\author}[1]{\oldauthor{#1}\newcommand{\theshortauthor}{#1}}
\newcommand{\shortauthor}[1]{\renewcommand{\theshortauthor}{#1}}
\cohead{\xrfill[0.525ex]{0.6pt}~\theshorttitle~\xrfill[0.525ex]{0.6pt}}
\cehead{\xrfill[0.525ex]{0.6pt}~\theshortauthor~\xrfill[0.525ex]{0.6pt}}
\newcommand{\theabstract}[1]{\par\bgroup\noindent\textbf{\textsf{Abstract.}} #1\egroup}
\newcommand{\thekeywords}[1]{\par\smallskip\bgroup\noindent\textbf{\textsf{Keywords.}}\newcommand{\and}{ $\bullet$ } #1\egroup}
\newcommand{\themsc}[1]{\par\smallskip\bgroup\noindent\textbf{\textsf{2020 Mathematics Subject Classification.}}\newcommand{\and}{ $\bullet$ } #1\egroup}
\newcommand*{\affilref}[1]{\ref{affiliation#1}}
\newcommand*{\affiliation}[3]{
	\footnotetext[#1]{\label{affiliation#2} #3}
}
\usepackage{siunitx} % for typesetting units

\numberwithin{equation}{section}
\numberwithin{figure}{section}
\numberwithin{table}{section}

\newcommand*{\defeq}{\coloneqq}

\renewcommand*{\leq}{\leqslant}

\newcommand*{\Reals}{\mathbb{R}}

\newcommand*{\negE}{\text{e}\mathord{-}}

\theoremstyle{definition}

\crefname{assumption}{Assumption}{Assumptions}
\Crefname{assumption}{Assumption}{Assumptions}

\title{Validation of the static forward Grad--Shafranov equilibrium solvers in FreeGSNKE and Fiesta using EFIT\texttt{++} reconstructions from MAST-U}
\shorttitle{FreeGSNKE and Fiesta static solver validation on MAST-U}
\author{
    K. Pentland\textsuperscript{\affilref{UKAEA}} 
    \and
    N. C. Amorisco\textsuperscript{\affilref{UKAEA}} 
    \and
    O. El-Zobaidi\textsuperscript{\affilref{UKAEA}}
    \and
    S. Etches\textsuperscript{\affilref{UKAEA}}
    \and
    A. Agnello\textsuperscript{\affilref{Hartree}}
    \and
    G. K. Holt\textsuperscript{\affilref{Hartree}}
    \and
    A. Ross\textsuperscript{\affilref{Hartree}}
    \and
    C. Vincent\textsuperscript{\affilref{UKAEA}}
    \and
    J. Buchanan\textsuperscript{\affilref{UKAEA}}
    \and
    S. J. P. Pamela\textsuperscript{\affilref{UKAEA}}
    \and
    G. McArdle\textsuperscript{\affilref{UKAEA}}
    \and
    L. Kogan\textsuperscript{\affilref{UKAEA}}
    \and
    G. Cunningham\textsuperscript{\affilref{UKAEA}}
}
\shortauthor{K.~Pentland et al.}
\date{\today}

\begin{document}
\maketitle

\affiliation{1}{UKAEA}{United Kingdom Atomic Energy Authority, Culham Campus, Abingdon, Oxfordshire, OX14 3DB, United Kingdom\newline (\email{kamran.pentland@ukaea.uk})}
\affiliation{2}{Hartree}{STFC Hartree Centre, Sci-Tech Daresbury, Keckwick Lane, Daresbury, Warrington, WA4 4AD, United Kingdom}

%%%%%%%%%%%%%%%%%%%%%%%%%%%%%%%%%%%%%%%%%%%%%%%%%%%%%%%%%%%%%%%%%
% Abstract

% • Context: give a brief background on the specific or general topic under study.
% • Problem/aims: state the main problem and the aims of your paper, i.e. how does your paper address the problem.
% • Methods/results: outline the methods you use (existing or novel) to solve the problem, stating the main results you obtain plus their significance.
% • Justify: detail why this problem is worth solving and how your approach is novel, given the context of what has been done in the field already.
    
\begin{abstract}\small
    \theabstract{%
    A key aspect in the modelling of magnetohydrodynamic (MHD) equilibria in tokamak devices is having access to fast, accurate, and stable numerical simulation methods. 
    There is an increasing demand for reliable methods that can be used to develop traditional or machine learning-based shape control feedback systems, optimise scenario designs, and integrate with other plasma edge or transport modelling codes.
    To handle such applications, these codes need to be flexible and, more importantly, they need to have been validated against both analytically known and real-world tokamak equilibria to ensure they are consistent and credible.
    In this paper, we are interested in solving the static forward Grad--Shafranov (GS) problem for free--boundary MHD equilibria. 
    Our focus is on the validation of the static forward solver in the Python-based equilibrium code \emph{FreeGSNKE} by solving equilibria from magnetics-only \emph{EFIT\texttt{++}} reconstructions of MAST-U shots.
    In addition, we also validate FreeGSNKE against equilibria simulated using the well-established MATLAB-based equilibrium code \emph{Fiesta}.
    To do this, we develop a computational pipeline that allows one to load the same (a)symmetric MAST-U machine description into each solver, specify the required inputs (active/passive conductor currents, plasma profiles and coefficients, etc.) from EFIT\texttt{++}, and solve the GS equation for all available time slices across a shot.
    For a number of different MAST-U shots, we demonstrate that both FreeGSNKE and Fiesta can successfully reproduce various poloidal flux quantities and shape targets (e.g.\ midplane radii, magnetic axes, separatrices, X-points, and strikepoints) in agreement with EFIT\texttt{++} calculations to a very high degree of accuracy.
    We also provide public access to the code/data required to load the MAST-U machine description in FreeGSNKE/Fiesta and reproduce the equilibria in the shots shown.
    }
    \thekeywords{%
        {MHD equilibria}%
        \and%
        {Grad--Shafranov}%
        \and%
        {FreeGSNKE}%
        \and%
        {Fiesta}%
        \and%
        {EFIT\texttt{++}}%
        \and%
        {MAST-U}%
    }
    % \themsc{
        % {65L70}%Error bounds for numerical methods for ordinary differential equations
        % \and%
        % {65Y05}%Parallel numerical computation
        % \and%
        % {65C99}%Probabilistic methods (in numerical analysis)
    % }
\end{abstract}

%%%%%%%%%%%%%%%%%%%%%%%%%%%%%%%%%%%%%%%%%%%%%%%%%%%%%%%%%%%%%%%%%
% Notes:

% \renewcommand\labelitemi{{\boldmath$\cdot$}}
% \renewcommand\labelitemi{$\vcenter{\hbox{\tiny$\bullet$}}$}

%%%%%%%%%%%%%%%%%%%%%%%%%%%%%%%%%%%%%%%%%%%%%%%%%%%%%%%%%%%%%%%%%
\section{Introduction} \label{sec:intro}

\subsection{Motivation and aims}

% motivation for the paper: MHD equilibrium usefulness/importance
Developing fast and accurate numerical methods for simulating the ideal magnetohydrodynamic (MHD) equilibrium of a magnetically-confined plasma is a crucial element in the design and operation of existing and future tokamak devices.
These solvers are used extensively to analyse different plasma scenarios, shapes, and stability, in addition to playing a critical role in the operation and optimisation of control and real-time feedback systems.

% explain the different types of solver
Many different equilibrium solvers have evolved over time and vary widely in terms of their design, purpose, ease-of-use, computational speed, and availability.  
For instance, they have been implemented in different programming languages and have harnessed various spatial discretisation schemes, techniques for solving nonlinear systems, and approaches for tackling optimisation problems.
They can, however, be broadly classified into two primary categories: \emph{static} and \emph{dynamic} (sometimes called \emph{evolutive}) solvers.

Our focus is on static solvers, which are time-independent and are designed to solve one of the following:
\begin{description}
    \item[``Forward problem'':] Solve for the plasma equilibrium using user-defined poloidal field coil currents, passive structure currents, and plasma current density profiles. 
    \item[``Inverse problem'':] Estimate poloidal field coil currents using user-defined constraints (e.g.\ isoflux and X-point locations) and plasma current density profiles for a desired plasma equilibrium shape.
    \item[``Reconstruction problem'':] Estimate (or fit) poloidal field coil currents and plasma current density profiles that yield a plasma equilibrium that ``best'' matches (noisy) measurement data from diagnostics around a tokamak (e.g.\ plasma/coil currents, magnetic readings, and density profiles). 
\end{description}
Dynamic ``forward'' and ``inverse'' solvers which tackle time-dependent equilibrium problems are also available. 
These solvers couple the static plasma equilibrium with time-evolving external conductor currents and voltages, however, they will not feature here.

% what are our aims?
In this paper, we focus on the \emph{free--boundary static forward MHD equilibrium problem}, which involves solving the Grad--Shafranov (GS) equation for a toroidally symmetric, plasma equilibrium.
As mentioned above, this requires user-defined poloidal field coil currents, passive structures currents (if available), and plasma current density profiles---see \cref{sec:forward_problem} for further details.
A vast array of numerical codes exist for solving this problem, however, our focus will be on two in particular: \emph{FreeGSNKE} and \emph{Fiesta} (more details to follow in \cref{sec:solvers}).
The aim of this work is to:
\begin{enumerate}[label=(\roman*)]
    \item validate that the static forward solver in FreeGSNKE can reproduce the equilibria obtained by a magnetics-only \emph{EFIT\texttt{++}} reconstruction on the MAST-U tokamak (in addition to the equilibria produced by Fiesta). 
    \item compare poloidal flux quantities, shape control measures (e.g.\ midplane radii, magnetic axes, separatrix positions, X-points, and strikepoints), and magnetics measurements from the solvers for a number of physically different MAST-U shots, using EFIT\texttt{++} as the reference solution. 
\end{enumerate}

% how do we do this
To enable an accurate and valid comparison of the results, we need to ensure that the static forward solvers in both FreeGSNKE and Fiesta are set up using the same set of \emph{input} quantities as \emph{output} by EFIT\texttt{++} (which itself solves the reconstruction mentioned before).
Firstly, this will require a consistent description of the MAST-U machine that includes the active coils, passive structures, wall/limiter, and magnetic diagnostics. 
We require details of their positions, orientations, windings, and polarity. 
Secondly, for each equilibrium, we need to appropriately assign values for both the active/passive conductor currents and plasma profiles parameters as calculated by EFIT\texttt{++}. 
In \cref{sec:inputs}, we provide more details on these input quantities and highlight differences between each code implementation.

% stress what we're actually doing
We should stress that while the reference EFIT\texttt{++} equilibria are obtained as equilibrium reconstructions, and are therefore the result of a fitting procedure from experimental measurements on the MAST-U tokamak, here we do not perform the same fitting procedure in FreeGSNKE or Fiesta.
We instead use the coil currents and plasma profiles parameters output by EFIT\texttt{++} as inputs to the static forward GS problems in FreeGSNKE and Fiesta.
Given a consistent set of inputs across all codes, we will demonstrate that all three return quantitatively equivalent equilibria. 

% why is this useful/why are we doing this?
Carrying out robust validation of static GS solvers\footnote{The validation of FreeGSNKE's \emph{dynamic} (evolutive) solver will not feature here and will be addressed in future work.}, against both analytic solutions and real-world tokamak plasmas, is critical for users that require consistent, and more importantly, trustworthy equilibrium calculations.
We carry out a rigorous comparison of the poloidal flux quantities, shape control targets, and magnetics measurements between each code, validating FreeGSNKE against both a forward (Fiesta) and reconstruction (EFIT\texttt{++}) equilibrium code.
Such in-depth and meticulous validation studies are rarely carried out for new or existing equilibrium solvers and as such we hope to make more consistent and quantitative validation possible by making the scripts required to do so publicly available (see end of paper for link to code and data).

\subsection{Related validation studies}
% overview
In this brief section, we will discuss a few validation efforts concerning static forward GS solvers, omitting inverse and reconstruction solvers here. 
While a comprehensive review of validation studies for all solver types, both static and dynamic (forward and inverse), would be worthwhile, it is far beyond the scope of the present work and warrants a dedicated future effort.

% discussion
We begin with \cite{hansen2023}, who demonstrate that the finite element-based static equilibrium solver within \emph{TokaMaker}, provides accurate solutions to the analytic ``Solov'ev'' and ``Spheromak'' fixed-boundary equilibrium problems. 
In addition, they simulate a SPARC equilibrium, comparing the last closed flux surface obtained to one from the inverse solver within \emph{FreeGS} (see \cref{sec:solvers} for more information).
Beyond this qualitative comparison, however, there is no in-depth quantitative validation of the accuracy of this flux surface or any other poloidal flux quantities associated with the SPARC equilibrium.
The all-purpose code \emph{NICE} is presented by \cite{faugeras2020} with a wide demonstration of its reconstruction and dynamic solver capabilities on WEST, TCV, and JET-like equilibria, however, the static solver appears to be untested (directly at least).
While \cite{jeon2015} demonstrate that the static forward solver within \emph{TES} can simulate equilibria for the KSTAR tokamak, their is no explicit investigation into the accuracy of the solutions obtained.

% discussion 
The lack of rigorous testing of static forward solvers highlights an urgent need for much more consistent benchmarking and validation for all different types of equilibrium codes that currently exist (or are yet to be developed).
The detailed cross-validation provided in this paper is rare due to the complexity of obtaining, setting up, and running different equilibrium codes with varying levels of documentation and access. 
We hope that the detailed description of our validation process and the availability of the corresponding data will enable easier validation of codes in the future.

\subsection{Paper structure}
% what do we do in this paper?
The rest of this paper will be structured as follows. 
In \cref{sec:solvers}, we provide an introduction to the three solvers FreeGSNKE, Fiesta, and EFIT\texttt{++}, briefly discussing their capabilities and prior usage in different areas of tokamak equilibrium modelling. 
In \cref{sec:forward_problem}, we outline the free--boundary static forward GS problem and note differences between the FreeGSNKE and Fiesta solution methods. 
Following this, we supply a more detailed description of the MAST-U machine and other more specific inputs required by each solver in \cref{sec:inputs}.
% probably a repetition?
% This includes encoding the MAST-U machine description (active coil, passive structure, limiter positions/windings etc.), assigning active coil and passive structure currents, and setting the form of the plasma density profiles correctly. 

In \cref{sec:numerics}, we present our numerical experiments, focusing on two different MAST-U shots, one featuring a conventional divertor configuration and the other a Super-X \citep{morris2014, morris2018}.
We begin by comparing FreeGSNKE and Fiesta, ensuring that we understand any key differences between the codes and how these differences may filter through when comparing with EFIT\texttt{++}. 
After this, we begin to assess the differences between equilibria (and other shape targets) from the solvers and those reconstructed from the diagnostics via EFIT\texttt{++}.
We find excellent agreement between all quantities assessed and highlight the accuracy of both FreeGSNKE and Fiesta. 
Finally in \cref{sec:discussion}, we discuss the implication of these results and close with a few suggestions for avenues of future work.

%%%%%%%%%%%%%%%%%%%%%%%%%%%%%%%%%%%%%%%%%%%%%%%%%%%%%%%%%%%%%%%%%
\section{The solvers} \label{sec:solvers}
In this section, we give a short introduction to the codes described in this paper and briefly discuss their capabilities. 

\subsection{FreeGSNKE}

% purpose/what it does
FreeGSNKE is a Python-based, finite difference, dynamic free--boundary toroidal plasma equilibrium solver developed by \cite{amorisco2024} and built as an extension of the publicly available \emph{FreeGS} code \citep{dudson2024}.
FreeGS features a Picard iteration-based static inverse solver for identifying the coil currents required to maintain different types of plasma configuration prior to experimentation---a brief introduction to constrained analysis can be found in \citet{jeon2015}[Sec. II.7.].
% , where user-specified controls (e.g.\ isoflux/X-point locations and coil currents) are used to determine a set of coil currents to achieve a desired equilibrium. 
% This type of analysis is used in 
% FreeGS also features a static forward solver, where coil currents are instead fixed by the user and the corresponding plasma equilibrium is calculated. 
% Both solvers in FreeGS use Picard iterations to solve the forward and inverse GS problems.
In conjunction with other equilibrium codes, FreeGS has been used extensively in recent years for the design of various tokamaks. % and have also been validated against other equilibrium codes. 
To our knowledge, FreeGS has aided the design of SPARC \citep{creely2020}, KSTAR \citep{lee2021}, WEST \citep{maquet2023}, Thailand Tokamak-1 \citep{sangaroon2023}, and MANTA \citep{mit2023}. 
It was also used in the design of ARC \citep{de_boucaud2022}, with further work on DIII-D EFIT reconstructed equilibria, and to design COMPASS-U \citep{kripner2018}, alongside Freebie \citep{artaud2012} and Fiesta, and to help develop the BLUEPRINT framework \citep{coleman2020}.

FreeGSNKE inherits the FreeGS inverse solver and introduces a static forward solver that uses a Newton--Krylov method (see e.g. \citet{Knoll04, carpanese2021}) to overcome the well-known numerical instability affecting Picard iterations.
Also introduced is a solver for the evolutive (dynamic) equilibrium problem, also based on the Newton--Krylov method. 
In the dynamic problem, Poynting's theorem is enforced on the plasma, coupling the circuit equations (that govern currents in the active coils/passive structures) and the GS equation itself \citep{amorisco2024}. 

% \textcolor{green}{FreeGSNKE relies on the original inverse solver part of the FreeGS suite to perform static constrained equilibrium analyses,} where user-specified controls (e.g.\ isoflux/X-point locations and coil currents) are used to constrain and determine a set of free (unconstrained) coil currents to achieve a desired equilibrium. 
% It is capable of solving the free--boundary static forward problem, where the plasma equilibrium is calculated by solving the GS equation with an appropriately constructed boundary condition. 
% Solving forward static GS problems requires a tokamak machine description, current values for both active coils and passive structures and appropriate plasma pressure and current density profile functions---more on this in \cref{sec:inputs}. 
% This problems are solved in a stable and efficient manner using a Newton--Krylov method, \textcolor{green}{which overcomes the well-known numerical instability affecting other equilibrium codes that rely on traditional Picard iterations.}  

% where have the code been used\validated? (on what machines?)
These features, as well as the widespread validation and use of the underlying FreeGS code, make FreeGSNKE a particularly versatile tool for studying the shape and control of plasma equilibria.
Its compatibility with other Python libraries, especially those with machine learning capabilities, facilitate its future development and integration with other plasma modelling codes.
For example, FreeGSNKE has been used to emulate scenario and control design in a MAST-U-like tokamak by \cite{agnello2024}, where their objective was to emulate flux quantities and shape targets (some of which we calculate here) based on a training library of input plasma profile parameters and active conductor currents.
% A Markov chain Monte Carlo algorithm was used for sampling and building this training set. 
% It is shown that the neural network emulators can achieve these tasks while being agnostic to the machine description and choice of nonlinear numerical solver, thereby avoiding the numerical issues and costs associated with solving the GS equation many times for many different plasma configurations.

% As mentioned before, FreeGS(NKE) has already been used with a simplified MAST(-U) machine description for constrained equilibrium analysis and emulation.
The static forward solver in FreeGSNKE has been validated against analytic solutions of the GS equation \citep{amorisco2024} and so now we go a step further by implementing the full MAST-U machine description and validate against EFIT\texttt{++} reconstructions.

\subsection{Fiesta}

% purpose/what it does
Fiesta is a free--boundary static equilibrium solver written in MATLAB and developed by \cite{cunningham2013}.
In addition to being able to carry out forward and inverse %(constrained)
equilibrium calculations, it is also capable of linearised dynamic modelling using the RZIp rigid plasma framework \citep{coutlis1999}.
% RZIp is a {\bb linearised?} plasma response model that measures how small changes in poloidal field coil voltages (via the circuit equations) affect plasma/conductor currents and the plasma radial/vertical position---useful for vertical displacement event modelling and the design of plasma control systems .
% {\bb Please correct me if I'm wrong and add more details as I really don't know much about Fiesta's use cases...}
% where have the code been used\validated? (on what machines?)
It has been used to inform design choices and carry out equilibrium analyses on a number of tokamaks including JET, DIII-D, NSTX, TCV, MAST(-U) \citep{windridge2011, cunningham2013}, MEDUSA-CR \citep{araya2021}, COMPASS-U \citep{vondracek2021}, SMART \citep{doyle2021, mancini2023}, STEP \citep{hudoba2023}, and EU-DEMO \citep{morris2021}. 

% why use Fiesta
Having already been used to simulate MAST(-U) and other tokamak equilibria, we run Fiesta alongside FreeGSNKE to demonstrate they both return quantitatively equivalent results given the same set of input data from EFIT\texttt{++}.
This cross-validation process should also help identify and explain any differences between the two different implementations.% and hopefully increase confidence in the output of both codes. 

\subsection{EFIT\texttt{++}}

% purpose/what it does
EFIT, first proposed by \cite{lao1985}, is a computational method for solving the reconstruction problem and is widely used as a first port of call for ``fitting'' plasma equilbria to measurement data from diagnostics within real-world tokamaks. 
These measurements come from diagnostics such as poloidal flux loops, pickup coils, Rogowski coils, motional Stark effect (MSE), and Thomson scattering systems, which are strategically located at key locations around the tokamak.
Written in Fortran, EFIT is used primarily for post-shot equilibrium reconstruction and has been implemented on a number of different tokamak devices (see below).
Our focus is on EFIT\texttt{++}, a substantial re-write in which the original EFIT code has been wrapped in a C\texttt{++} driver to handle data flow, which in turn is wrapped in a highly configurable Python layer for input and output checking. 
It is currently in use on the MAST-U tokamak \citep{appel2006} and was previously deployed on JET \citep{appel2018}.
We note that EFIT\texttt{++} is run routinely for all MAST-U plasma shots using magnetic diagnostic data only and, if available, MSE data to improve the accuracy of core profiles.
In addition to this, EFIT\texttt{++} is also set up to use Thomson scattering data if required \citep{berkery2021, kogan2022}.

% Our focus is on EFIT\texttt{++}, a substantial re-write of the original EFIT code that includes a C\texttt{++} driver for handling data flow. 

To solve the inverse problem, EFIT\texttt{++} requires descriptions of the plasma pressure and toroidal current profiles which are typically expressed using basis functions (whose coefficients are to be adjusted during the fitting process).
Next, the linearised GS equation is solved using an initial guess for the poloidal flux. 
The feasibility of the calculated flux with respect to the diagnostic measurement data is then measured by solving a linearised least-squares minimisation problem. 
During this process, the variable parameters such as the conductor currents and profile coefficients are adjusted to improve the fit. 
This iterative process repeats until the conductor currents, profile coefficients, and poloidal flux, together, return a valid solution to the GS equation at the required tolerance.
For more technical details, refer to \cite{lao2005}, \cite{appel2018}, and \cite{bao2023}.

% where have the code been used\validated? (on what machines?)
Different versions of EFIT, each with their own configurations and modifications, have been used for equilibrium reconstruction on a vast array of tokamak devices. 
Without providing an exhaustive list, it has been deployed on JET \citep{appel2018}, MAST(-U) \citep{kogan2022}, EAST \citep{bao2023}, DIII-D \citep{lao2005}, START \citep{appel2001}, KSTAR \citep{lee1999}, NSTX \citep{sabbagh2001}, and ITER \citep{lao2022}.
Given its history of widespread use on many different tokamak devices, we use EFIT\texttt{++} as a source of trusted reference equilibria, against which to compare those produced by FreeGSNKE and Fiesta.

% While this validation process will involve comparing the equilibria (and other shape quantities) with those from EFIT\texttt{++}, we should note that any \emph{errors} we show the codes are not really errors but 

%  % MAST-U overview
% Aims: to explore more compact (spherical) tokamak designs that can achieve more efficient plasma performance and test alternate reactor technologies for future fusion devices such as ITER, STEP, and ...?
% Increased pulse length, heating power, currents, magnetic fields. 
% In particular, MAST-U is experimenting with the ``Super-X'' divertor and can therefore operate with different divertor configurations that will be needed to handle the extreme heat and power loads of (larger) future devices. 

%%%%%%%%%%%%%%%%%%%%%%%%%%%%%%%%%%%%%%%%%%%%%%%%%%%%%%%%%%%%%%%%%
\section{The static forward Grad--Shafranov problem} \label{sec:forward_problem}
In this paper, we are interested in solving the GS equation
\begin{align} \label{eq:Grad--Shafranov}
    \Delta^{*} \psi = -\mu_0 R \underbrace{\left( J_{p} + J_{c} \right)}_{= J_{\phi}}, \quad (R,Z) \in \Omega,
\end{align}
in the cylindrical coordinate system $(R, \phi, Z)$ for the poloidal flux $\psi(R,Z)$\footnote{Note that some numerical solvers (e.g.\ Fiesta) define $\psi$ using the Weber whereas some (e.g.\ FreeGSNKE and EFIT\texttt{++}) define it using the Weber/$2\pi$. } \citep{grad1958, shafranov1958}.
This equation describes the equilibrium of a magnetically confined plasma in which the plasma pressure and magnetic forces acting upon it are in balance.   
It can be derived by exploiting toroidal symmetry in the ideal MHD equations (see \citet{jardin2010}[Chp. 4]).

In \eqref{eq:Grad--Shafranov}, $\mu_0$ represents magnetic permeability in a vacuum and $\Delta^* \defeq R \partial_R R^{-1} \partial_R + \partial_{ZZ}$ is a linear elliptic operator.
The toroidal current density $J_{\phi}(\psi, R, Z) \defeq J_{p}(\psi, R, Z) + J_{c}(R, Z)$ contains a contribution from both the plasma $J_{p}$ and any toroidally symmetric conducting metal structures external to the plasma $J_{c}$ (e.g.\ active poloidal field coils and passive structures around the tokamak).
The total poloidal flux $\psi \defeq \psi_p + \psi_c$ is also made up of a plasma $\psi_p$ and external conductor $\psi_c$ contribution. 
We wish to solve \eqref{eq:Grad--Shafranov} over a two-dimensional computational domain $\Omega \defeq \Omega_{p} \cup \Omega_{p}'$ where $\Omega_p$ represents the plasma region\footnote{The boundary of $\Omega_p$ is defined as the closed $(R,Z)$ contour in $\Omega$ that passes through the X-point closest to the magnetic axis (see closed red contour in \cref{fig:MASTU}).} and $\Omega_{p}'$ is its complement. 

The plasma current density, non-zero only within the plasma region $\Omega_{p}$, takes the form
\begin{align} \label{eq:plasma_currents}
     J_{p}(\psi, R, Z) = R \frac{\mathrm{d} p}{\mathrm{d} \psi} + \frac{1}{\mu_0 R} F \frac{\mathrm{d} F}{\mathrm{d} \psi}, \quad (R,Z) \in \Omega_{p},
\end{align}
where $p \defeq p(\psi)$ is the isotropic plasma pressure profile and $F \defeq F(\psi) = R B_{\phi}$ is the toroidal magnetic field profile ($B_{\phi}$ is the toroidal component of the magnetic field).
The particular choice of profile functions used in $J_p$ will be discussed in \cref{sec:inputs}.
The current density generated by $N_{c}$ external conductors is given by
\begin{align} \label{eq:ext_currents}
     J_{c}(R, Z) &= \sum_{j=1}^{N_{c}} \frac{I_j^{c}(R,Z)}{A_j^c}, \quad (R,Z) \in \Omega, \\
     I_j^c(R,Z) &= \begin{cases}
    I_j^c & \ \text{if} \ (R,Z) \in \Omega_j^c, \vspace*{0.00cm} \\ 
    0 & \ \text{elsewhere}, \vspace*{0.00cm} \nonumber
\end{cases}
\end{align}
where $\Omega_j^c$, $I_j^{c}$, and $A_j^c$ are the domain region, current, and cross-sectional area of the $j$\textsuperscript{th} conductor, respectively.
Note that external conductors can lie inside $\Omega$ as well as outside of it.
% {\bb I think we should keep this as they're define terms in \eqref{eq:BC}. I also found it really helpful to know these terms when first learning about the forward problem. }

To complete the free--boundary problem, an appropriate Dirichlet boundary condition must also be specified on the domain boundary $\partial \Omega$---which we discuss in the next section. 
The dependence of $J_{p}$ on $\psi$ makes \eqref{eq:Grad--Shafranov} a nonlinear elliptic partial differential equation.
% \begin{align}
%     J_c(r, z) = \sum_{j=1}^{N_c} \frac{I_j^c}{A_j^c} \int_{\Omega_j^c} \delta(r - r') \delta(z - z') \, dr' \, dz'.
% \end{align}

\subsection{Solving the problem}
Here, we briefly outline the steps typically carried out when numerically\footnote{Analytic solutions to the GS problem do exist in limited cases---see \citet{cerfon2010} for some examples.} solving the static free--boundary (forward) GS problem \citep{jardin2010, jeon2015}.
For more specific details on how each of the solvers do this in practice, we refer the reader to the respective code documentation.

Before solving, we assume that a number of input parameters have already been provided by the user including: a machine (tokamak) description, conducting structure (active coil and passive structure) currents, and plasma profile functions (and parameters). 
More details on the specific inputs required for generating free--boundary equilibria on MAST-U with each of the codes will be described in \cref{sec:inputs}.

\subsubsection*{Step one}
Denote the total flux by $\psi^{(n)}(R,Z)$, where $n = 0,1,\ldots$ is the iteration number, and generate an appropriate guess $\psi^{(0)}$ to initialise the solver\footnote{Note that $\psi_c^{(n)}$ is known exactly (it is given by the second term in \eqref{eq:BC}) and so we only require an initial guess for $\psi^{(0)}_p$.}.

\subsubsection*{Step two}
Calculate the values of the flux on the computational boundary $\partial \Omega$ (i.e.\ the Dirichlet boundary condition) using 
\begin{align} \label{eq:BC}
    \psi^{(n)} \bigg\rvert_{\partial \Omega}
    &= \int_{\Omega_{p}} G(R,Z;R',Z') J_{p}(\psi^{(n)}, R',Z') \ \mathrm{d}R' \mathrm{d}Z' \nonumber \\
    \quad &+ \sum_{j=1}^{N_{c}} \frac{1}{A_j^c} \int_{\Omega_j^c} G(R,Z;R',Z') I_j^{c}(R',Z') \ \mathrm{d}R' \mathrm{d}Z', % \sum_{j=1}^{N_{c}} G(r,z;r_j^{c},z_j^{c}) I_j^{c},
\end{align}
where the first and second terms are the contributions from $\psi^{(n)}_{p}$ and $\psi^{(n)}_{c}$ on the boundary, respectively.
$G$ is a Green's function for the operator $\Delta^*$ containing elliptic integrals of the first and second kind---it can be calculated by solving \eqref{eq:Grad--Shafranov} with $\psi_c$ alone (see \citet{jardin2010}[Chp. 4.6.3]). 
To calculate \eqref{eq:BC}, the plasma domain $\Omega_p$ (i.e.\ the area contained within the last closed flux surface) needs to be identified---see \citet{jeon2015}[Sec. 5] for how to do this.
Once found, the integral itself can be calculated a number of different ways, for example, using von Hagenow's method \citep{jardin2010}[Chp. 4.6.4].
% \footnote{The second term is found by integrating the product of the Green's function with \eqref{eq:ext_currents} over $\Omega$.}

\subsubsection*{Step three}
To solve the nonlinear problem, both EFIT\texttt{++} and Fiesta use Picard iterations \citep{kelley1995}, where the $n$-th iteration consists of calculating the total flux $\psi^{(n+1)}$ according to
\begin{align} \label{eq:nonlinear_picard}
    \Delta^* \psi^{(n+1)} = -\mu_0 R J_{\phi}(\psi^{(n)}, R, Z), \quad (R,Z) \in \Omega,
\end{align}
together with boundary condition \eqref{eq:BC}.
In a finite difference implementation this requires spatially discretising the elliptic operator $\Delta^*$.
For example, FreeGSNKE uses fourth-order accurate finite differences while Fiesta uses a second-order accurate (fast) discrete sine transform.

\subsubsection*{Step four}
Check whether or not the solution meets a pre-specified tolerance, e.g.\ a relative difference such as
\begin{align} \label{eq:tolerance}
    \frac{ \max | \psi^{(n+1)} - \psi^{(n)} |}{\max(\psi^{(n)}) - \min(\psi^{(n)})} < \varepsilon.
\end{align}
If so, we stop the iterations, otherwise we continue.

Both FreeGSNKE and Fiesta are set to use the same relative tolerance $\varepsilon = 1\negE6$ and while FreeGSNKE uses the criterion in \eqref{eq:tolerance}, we should note that Fiesta uses a slightly different relative criterion based on values of $J_p$ at successive iterations instead of $\psi$.
This should make little difference to the comparison. 

\subsubsection*{Comments}
Picard iterations are very effective at tackling inverse GS problems, which is the primary use case for EFIT\texttt{++} and Fiesta. 
However, it is well-known that these iterations are unstable when applied to forward GS problems.
This manifests itself in the form of vertically unstable equilbria that artificially move between successive Picard iterations \citep{carpanese2021}.
This arises as a result of a combination of known physical instabilities in highly-elongated plasmas and mathematical features of the Picard method itself (which stem from a combination of steep gradients in the nonlinear function and a poor initial guess to the solution).
Newton-based methods can overcome this instability (see e.g. \citet{carpanese2021}). FreeGSNKE implements a Jacobian-free Newton--Krylov method (see \citet{amorisco2024}[App. 1] for further details). 
This is used to solve directly for the roots, $\psi$, of
\begin{align} \label{eq:nonlinear_root}
    \Delta^* \psi + \mu_0 R J_{\phi}(\psi, R, Z) = 0, \quad (R, Z) \in \Omega.
\end{align}
As with the Picard iteration, solving this problem still requires an appropriate initial guess and the calculation of the (nonlinear) boundary condition \eqref{eq:BC}.
% We can come back to this a bit when comparing the times to solution.
% The NK method has been shown by \cite{amorisco2024} to be much more numerically stable for both static and dynamic GS problems.
% {\bb How much detail do we want to go into?...}

%%%%%%%%%%%%%%%%%%%%%%%%%%%%%%%%%%%%%%%%%%%%%%%%%%%%%%%%%%%%%%%%%
\section{Input parameters (for MAST-U)} \label{sec:inputs}

To solve the forward problem, we need to ensure that the inputs to both FreeGSNKE and Fiesta are consistent with those used %(or ``fit'') 
by EFIT\texttt{++} on MAST-U.
We require:
\begin{enumerate}
    \item an accurate and representative MAST-U machine description containing the:
    \begin{enumerate}
        \item active poloidal field coils.
        \item passive structures.
        \item limiter/wall structure.    
        \item poloidal fluxloops and magnetic pickup coils.
    \end{enumerate}
    \item the fitted values of the coil currents of both active coil and passive structures.
    \item the functional form chosen for the plasma profile functions and the corresponding fitted parameter values.
    \item any additional parameters specific to either FreeGSNKE or Fiesta.
\end{enumerate}

In the following sections, we outline how these inputs are configured for each of the codes.
We note that the fluxloops and magnetic pickup coils are \emph{not required} to solve the forward problem, but rather are used during the post-simulation analysis in \cref{sec:numerics}.
% and note that the MAST-U machine description and all other input/configuration data can be found in the public repository: \url{https://github.com/...}.

\subsection{Machine description}

The following machine description had already been implemented in both EFIT\texttt{++} and Fiesta and has now been set up in FreeGSNKE.
We note here that numerical experiments (in \cref{sec:numerics}) across all codes are simulated on a $65 \times 65$ computational grid on $\Omega = [0.06, 2.0] \times [-2.2, 2.2]$, as this is the resolution EFIT\texttt{++} is run at during MAST-U reconstructions.

\subsubsection{Active coils}
\begin{figure}[t!]
    \centering
    \includegraphics[width=0.49\textwidth]{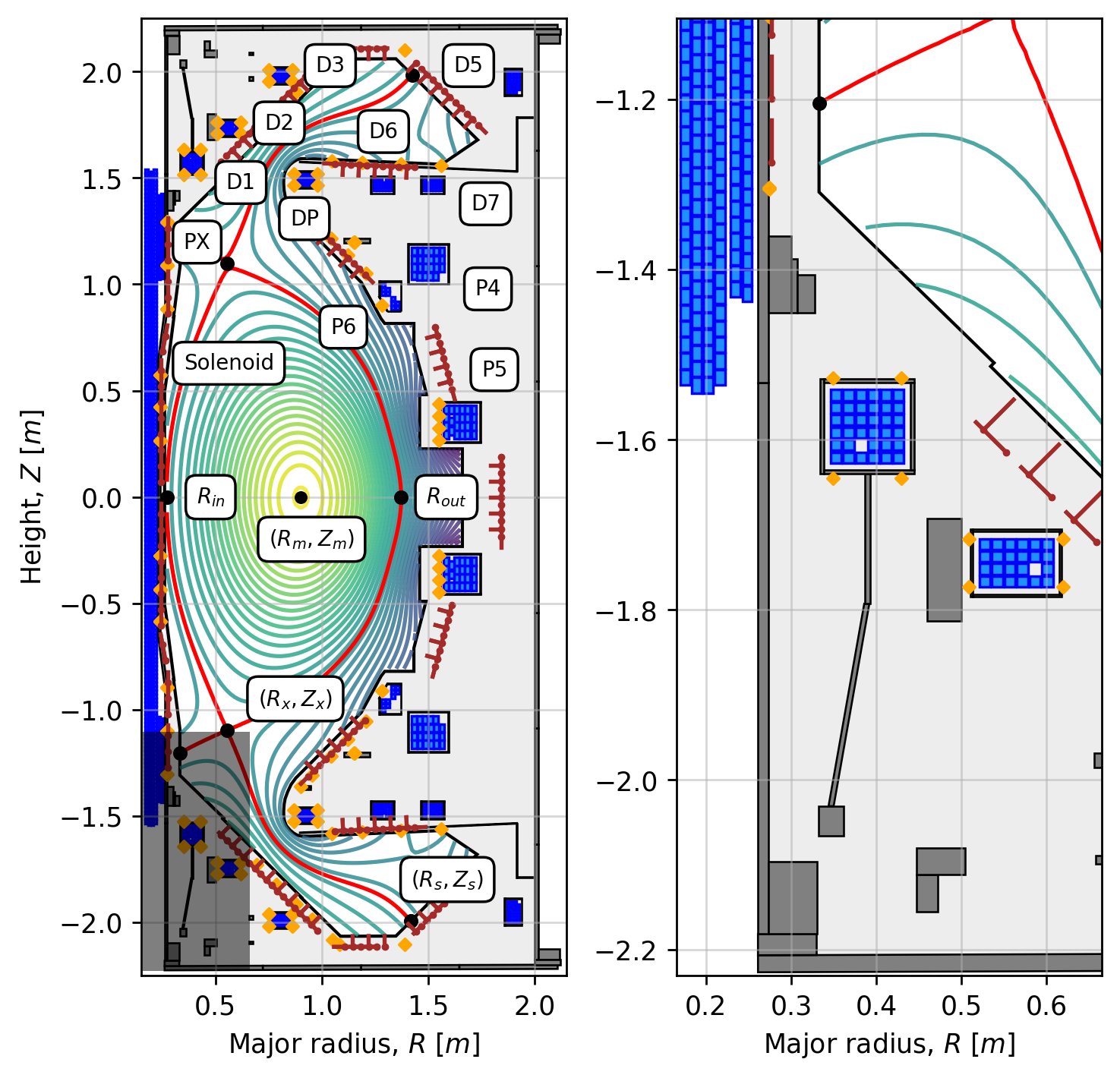}
    \caption{Left: poloidal cross-section of the MAST-U machine (as used across all three codes) with FreeGSNKE-simulated $\psi$ contours of shot $45292$ ($t=0.55$s). 
    Shown are the 12 active coils (blue), the passive structures (dark grey), the limiter/wall (black line, white interior), and the locations of shape targets to be tracked in experiments later on.
    Also shown are locations/orientations of magnetic diagnostics, including the flux loops (orange diamonds) and the pickup coils (brown dots/lines).
    Right: magnified view of the lower left corner of the machine (dark shaded box on left figure), showing individual filaments/windings inside some of the active coils and passive structures.
    }
    \label{fig:MASTU}
\end{figure}

% general details
MAST-U contains 12 active poloidal field coils whose voltages can be varied for shaping and controlling the plasma \citep{ryan2023}. 
In \cref{fig:MASTU}, we display a poloidal cross-section of the machine (as is implemented in all three codes) with an example equilibrium from a MAST-U shot (all simulated and plotted using FreeGSNKE). 
The solenoid, named P1 on MAST-U, generates plasma current and a poloidal magnetic field while P4/P5/PC (the latter of which is not currently connected to the machine) are used for core radial position and shape control.
P6 is used for core vertical control, D1/D2/D3/PX for X-point positioning and divertor leg control, and DP for further X-point positioning and flux expansion. 
Coils D5 and D6/D7 are used for Super-X leg radius and expansion control, respectively.

% code specifics
All active coils (except the solenoid) have an upper (labelled in \cref{fig:MASTU}) and lower component (not labelled) that are wired together in the same circuit.
All upper and lower coils are wired in series, except for the P6 coil, whose upper and lower components are connected in anti-series so that it can be used for vertical plasma control.
Each coil consists of a number of filaments/windings (plotted as small blue rectangles on the right hand side of \cref{fig:MASTU}) each with their own central position $(R,Z)$, width and height ($dR,dZ)$, polarity ($+1$ in series, $-1$ in anti-series), and current multiplier factor (used for the solenoid only). 
For the scope of the poloidal field, individual windings are modelled as infinitesimally thin toroidal filaments in both FreeGSNKE and Fiesta. 
Each filament also features its own resistivity value, however, this is not used here where we only deal with static equilibria.
% Specific details for each are stored in the aforementioned code repository.

% note on asymmetry
We should note that when EFIT\texttt{++} fits the active coil currents to diagnostic data, it \emph{does not} treat the upper and lower windings of the same active coil as being linked in series. 
Instead, current values in the upper and lower windings are measured using independent Rogowski coils\footnote{The active coil currents are approximated using the difference between measured internal (coil only) and external (coil plus coil case) Rogowski coil currents \citep{ryan2023} and are then fit (alongside all other quantities of interest). The same process is used to fit coil case currents---see \cref{sec:assign_currents}.} and are therefore fit to slightly different values.
The relative difference between the upper/lower coil current values is very small and so using this configuration makes little difference to equilibria generated by EFIT\texttt{++}.
We refer to this configuration as having \emph{asymmetric} (or \emph{up-down independent}) coil currents.
% This is to allow for more freedom in the least-squares optimisation, thereby increasing the flexibility of the model and hence the chance of a successful equilibrium reconstruction.
% and therefore ensure that EFIT\texttt{++} is not over-constrained {\bb (is this correct?)}. This restriction is currently under investigation. 
For both FreeGSNKE and Fiesta, we have the option to model the pairs of active coils as either symmetric (connected in series/anti-series as they are in the real MAST-U machine) or asymmetric (as in EFIT\texttt{++}).
All of the experiments presented in \cref{sec:numerics} are carried out using the asymmetric coil setup so that we can recreate the EFIT\texttt{++} configuration as closely as possible.

\subsubsection{Passive structures}

% why are passive structures important to include in static recontrcutions?
Both the active coils and the plasma itself induce significant eddy currents in the toroidally continuous conducting structures within MAST-U \citep{mcardle2008, berkery2021}.
This is especially the case in spherical tokamak devices, due to the close proximity of passive structures to the plasma core and active coils.
These currents significantly impact the plasma shape and position, making their inclusion in the modelling process essential to obtaining accurate equilibrium simulations.

% MAST-U
% I would not use "filaments" here
The complete MAST-U machine description includes a total of 150 passive structures, making up the vessel, centre column, support structures, gas baffles, coil cases, etc. 
This number excludes a few structures that are not included in the EFIT\texttt{++} model. 
These include the graphite tiles (which do not carry much current) and the cryopump (which contains a toroidal break to prevent large toroidal currents flowing around the machine).
Each passive structure is represented by a parallelogram in the poloidal plane, defined by its central position $(R,Z)$, width and height $(dR,dZ)$ and two angles, $(\theta_1, \theta_2)$\footnote{$\theta_1$ is the angle between the horizontal and the base edge of the parallelogram while $\theta_2$ is the angle between the horizontal and right hand edge (i.e.\ $\theta_1 = \theta_2 = 0$ defines a rectangle).}.
Such parallelograms can be seen on the right hand side of \cref{fig:MASTU}.
Given that all three codes require these parallelogram structures to simulate equilibria in the $(R,Z)$ plane, we should note that this passive structure model is a reduced axisymmetric representation of the true three-dimensional MAST-U vessel which contains toroidal breaks for vessel ports among other things---see \citet{berkery2021}[Fig. 12] for a depiction of the full 3D model.

Both FreeGSNKE and Fiesta model the poloidal field associated with each passive structure by uniformly distributing its current density over the poloidal cross-section. 
This can be done by ``refining'' (i.e.\ subdividing) each passive structure into individual filaments.
We revisit this in \cref{sec:assign_currents} when we discuss how to assign currents to the passive structures.
% to enable a more accurate and realistic distribution of current for structures that may have a (relatively) large surface area. This simply requires taking the parallelogram shape of the passive filament and partitioning it into a number of sub-filaments, using a given area and length scale. 

\subsubsection{Limiter/wall structure}
The purpose of the limiter/wall in FreeGSNKE and Fiesta is to confine the boundary of the plasma. 
In all three codes it is described by 98 pairs of $(R,Z)$ coordinates that form the closed polygonal shape seen in \cref{fig:MASTU} (enclosing the flux contours).
The plasma core is forced to reside within the limiter region, with the last closed flux surface being either fully contained in this region or tangent to its polygonal edge.
The limiter and the wall are taken to be the same. 

\subsubsection{Fluxloops and magnetic pickup coils}

% general details
Magnetic diagnostics are crucial for stable plasma control and equilibrium reconstruction, among other tasks. 
Although MAST-U is equipped with several different types of magnetic diagnostics such as Rogowski, saddle, and Mirnov coils, here we focus on the setup of the flux loops and magnetic pickup coils within FreeGSNKE and EFIT\texttt{++}.
For a comprehensive overview of MAST-U's magnetic diagnostics, refer to \cite{ryan2023}.

% poloidal fluxloops
A fluxloop in MAST-U is a copper cable wound in a single toroidal loop, located at a fixed poloidal location $(R,Z)$. 
There are currently 102 fluxloops installed at various poloidal locations on MAST-U where they are used to directly measure the $\psi(R,Z)$ at their specific locations (represented by the orange diamonds in \cref{fig:MASTU}). 

% pickup coils
The magnetic pickup coils are multi-turn copper coils that measure the strength of the magnetic field (in Tesla) orthogonal to its orientation.
MAST-U contains $354$ pickup coils, each positioned at a fixed location\footnote{Due to assumed toroidal symmetry, we effectively ignore $\phi$ here.} $(R, \phi, Z)$ with (normalised) orientation $\bm{\hat{n}} = (\hat{R}, \hat{\phi}, \hat{Z})$.
The magnetic field strength can be determined by calculating $\bm{B} \cdot \bm{\hat{n}}$, where
\begin{align*}
    \bm{B}(R,\phi,Z) = (B_R, B_{\phi}, B_Z) = \left( -\frac{1}{R} \frac{\mathrm{d} \psi}{\mathrm{d} Z}, \frac{F(\psi)}{R}, \frac{1}{R} \frac{\mathrm{d} \psi}{\mathrm{d} R} \right).
\end{align*}
The pickup coils are represented by the brown dots and lines (indicating the locations and orientations of the coils, respectively) in \cref{fig:MASTU}.

% comments
The simulated readings from these two different sets of diagnostics can be straightforwardly calculated (\emph{after} an equilibrium has been simulated) in FreeGSNKE.
We stress that experimental diagnostic measurements from MAST-U are not required to solve the static forward GS problem.
Later in \cref{sec:numerics}, we will compare the simulated readings from FreeGSNKE with the measurement data from MAST-U (i.e.\ data used to carry out the EFIT\texttt{++} reconstruction) over the course of a shot to further validate the accuracy of the FreeGSNKE simulations. 

\subsection{Assigning currents} \label{sec:assign_currents}

\subsubsection{Active coils}

% assigning the currents
As mentioned before, both FreeGSNKE and Fiesta have the option to use up-down symmetric or asymmetric (independent) active coil current assignments. 
To set the coil currents in the asymmetric setting, we assign the individually calculated upper/lower coil currents from EFIT\texttt{++} directly to the corresponding coils in FreeGSNKE and Fiesta without modification.
If we were to use the symmetric coil setting, however, each of the 12 active coils in FreeGSNKE and Fiesta require a single current value. 
To set each one, we could, for example, take the average of the corresponding upper and lower coil currents from EFIT\texttt{++}, making sure that the correct polarity of each current is also assigned.

\subsubsection{Passive structures}

% assigning the currents
Due to the different ways they are modelled in EFIT\texttt{++}, care needs to be taken when assigning the fitted passive structure currents to the 150 structures defined in Fiesta and FreeGSNKE.
For example, current values for each coil case are fit by EFIT\texttt{++} explicitly (using the Rogowski coil measurements mentioned before), which makes it easy to assign them directly in both FreeGSNKE and Fiesta.
% These currents are available for each coil case being modelled and can be set directly in FreeGSNKE and Fiesta. 
Other passive structure currents in the vessel, centre column, gas baffles, and support structures, are not, however, measured (and therefore fit) directly.
To reduce the degrees of freedom in EFIT\texttt{++}, these passive structures are modelled in groups, each referring to a single current value\footnote{An electromagnetic induction model is used to calculate current values to adopt as priors in the fit---refer to \cite{mcardle2008} and \cite{berkery2021} for more details.}, thereby reducing the computational runtime and avoiding some issues created by having too much freedom in the distribution of current around the machine \citep{berkery2021}[Sec. 3].
In total, in MAST-U there are 20 groups: 14 for the vacuum vessel, 2 for the gas baffles, 2 for passive stabilisation plates, and 2 for divertor coil supports \citep{kogan2021}.
To set the correct current for each structure in a group, we follow \citet{berkery2021}[Sec. 3] and distribute the group current proportionally to each structure based on its fraction of the total cross-sectional area within the group.

% refining the strucures
As mentioned before, both FreeGSNKE and Fiesta have the option to ``refine'' the 150 passive structures into sets of filaments for improved electromagnetic modelling. 
This involves taking each parallelogram structure, dividing its area (or length) into filaments of approximately the same size, and then evenly distributing the structure current amongst them uniformly.
The density of such filaments over the poloidal section of each structure can be adjusted as desired in FreeGSNKE and Fiesta\footnote{
In our numerical experiments, FreeGSNKE uses $7,297$ refined filaments with cross-sectional areas ranging from $0.03$cm\textsuperscript{2} to $2.4$cm\textsuperscript{2} (median area is $0.13$cm\textsuperscript{2}).
In Fiesta, the refinement is carried out slightly differently and uses $7,030$ refined filaments, with areas ranging from $0.16$cm\textsuperscript{2} to $0.62$cm\textsuperscript{2} (median $0.24$cm\textsuperscript{2}).}.

\subsection{Profile functions}

% explain Lao profile
To complete the set of input parameters for the static forward GS problem we need consistent plasma current density profile functions across the codes. 
In a magnetics-only EFIT\texttt{++} reconstruction on MAST-U \citep{appel2018,kogan2022}, the pressure and toroidal current profiles in \eqref{eq:plasma_currents} are defined using the following polynomials, sometimes referred to as the ``Lao profiles'' as first introduced by \cite{lao1985} in the original EFIT code:
\begin{align}
    \frac{\mathrm{d} p}{\mathrm{d} \tilde{\psi}} = \sum_{i=0}^{n_p} \alpha_i \tilde{\psi}^i - \bar{\alpha} \tilde{\psi}^{n_p + 1} \sum_{i=0}^{n_p} \alpha_i, \\
    % \quad \text{and} \quad 
    F \frac{\mathrm{d} F}{\mathrm{d} \tilde{\psi}} = \sum_{i=0}^{n_F} \beta_i \tilde{\psi}^i - \bar{\beta} \tilde{\psi}^{n_F + 1} \sum_{i=0}^{n_F} \beta_i,
\end{align}
with coefficients $\alpha_i, \beta_i \in \Reals$.
Note here that 
\begin{align}
    \tilde{\psi} = \frac{\psi - \psi_a}{\psi_b - \psi_a} \in [0,1],
\end{align}
is the normalised poloidal flux where $\psi_a = \psi(R_m, Z_m)$ and $\psi_b = \psi(R_X, Z_X)$ are the values of the flux on the magnetic axis and plasma boundary\footnote{In non-diverted (limited) plasmas, the boundary flux value is instead defined where the plasma contacts the limiter.}, respectively.

To avoid over-fitting and solution degeneracy problems, EFIT\texttt{++} uses lower-order polynomials ($n_p = n_F = 1$) for the magnetics-only reconstructions we validate against in \cref{sec:numerics}.
The logical parameters $\bar{\alpha} = \bar{\beta} = 1$ are set to enforce homogeneous Dirichlet boundary conditions on the plasma boundary (i.e.\ $p'(\tilde{\psi}=1) = FF'(\tilde{\psi}=1) = 0$).
Neumman boundary conditions (on the profile derivatives) can also be enforced if required---see \cite{berkery2021}[Sec. 2]. %{\bb Are they enforced in magnetics-only EFIT\texttt{++}? I don't think so...}

We therefore assign values of the coefficients $\alpha_i$ and $\beta_i$ as determined by EFIT\texttt{++} and proceed to normalise the profile functions using the value of the total plasma current $I_p$ fitted by  EFIT\texttt{++}. 
This step is nominally redundant but represents an additional check that ensures the profile functions set in FreeGSNKE and Fiesta are exactly the same as in EFIT\texttt{++}.

For EFIT\texttt{++} reconstructions that use data from both the magnetics and the MSE diagnostics, \emph{tension spline} representations of the profiles are used---see \cref{app:tension_spline}.
Although FreeGSNKE can indeed simulate equilibria using spline-based profiles (and a number of other commonly used profiles), we do not present those results here as they are of a comparable accuracy to the magnetics-only results presented in \cref{sec:numerics}.
Additionally, the spline profiles were unavailable in Fiesta for comparison.

\subsection{Other parameters and code specifics}
Here, we detail a few other parameters that need to be set in order to run the forward solvers in both Fiesta and FreeGSNKE. 

Firstly, in Fiesta, we need to specify a \texttt{feedback} object that mitigates the vertical instability that manifests itself via the Picard solver.
One option (via the \texttt{feedback2} object) is to monitor the vertical plasma position during the solve and modify the P6 coil current(s) (i.e.\ the radial field it produces) to correct the position error. 
Modifying the P6 coil current would, however, defy the purpose of the comparison with the EFIT\texttt{++} equilibria.
To keep the P6 coil current(s) (and all other inputs currents fixed), we therefore opt to use the \texttt{feedback3} object instead---this uses a variation of the method presented by \cite{yoshida1986}.
This object introduces a second (outer) nonlinear solver loop.
The inner loop solves for the equilibrium (via the Picard iterations) with respect to a specified magnetic axis location by adding `synthetic' radial and vertical magnetic fields.  
The outer loop then minimises these synthetic fields using a gradient search method, returning an optimal solution for the magnetic axis position, and therefore the equilibrium.
This additional outer loop drastically slows Fiesta compared to the \texttt{feedback2} setting, however, we believe that it is necessary for a direct comparison with FreeGSNKE and EFIT\texttt{++}. 

% This object stabilises the plasma vertical position by constraining the flux difference between two vertically separated points to a specified value (typically zero).
% This is done by adjusting the radial field produced by the P6 coils to cancel any error at each step within the Picard solver (the same process exists in EFIT\texttt{++}).
% By scanning over a range of vertical positions, the appropriate flux difference to give a specified P6 current (i.e.\ the input from EFIT\texttt{++}) can be determined using Newton's method. 
% This additional nonlinear solver loop drastically slows Fiesta compared to the \texttt{feedback2} setting, however, we believe that it is necessary for a direct comparison with FreeGSNKE and EFIT\texttt{++}. 

% This estimates the vertical plasma position by iteratively solving (using a Newton-based method) a nonlinear relationship between the fixed P6 current and the flux difference between two vertically separated locations. 
% This vertical position is then stabilised using an auxiliary radial field.

Secondly, both Fiesta and FreeGSNKE require a prescription for the toroidal field. 
This is provided through a parameter called \texttt{irod} in Fiesta, specifying the total current in the central toroidal field conductor bundle, and through the value of $f_{\mathrm{vac}} \defeq RB_{tor}$ in FreeGSNKE. 
We note, however, that the toroidal field does not affect the equilibrium calculations themselves. Calculation, for example, of safety factors and beta values would be affected, but we do not consider them here. 
For completeness, in FreeGSNKE we set $f_{\mathrm{vac}}$ using the EFIT\texttt{++} sourced value, and in Fiesta we use \texttt{irod} $= 5e6f_{\mathrm{vac}}$.

Finally, for the simulations in \cref{sec:numerics}, Fiesta occasionally struggles to converge for a number of time slices in the two shots shown. 
This could be due to a combination of the nonlinearity of the GS equation, a poor initial guess for $\psi_{p}$ (or $J_p$), and perhaps the instability of the Picard solver. 
To remedy this, the results we present here are obtained by providing Fiesta with the $J_p$ field calculated by EFIT\texttt{++}, as an initial guess in the Fiesta forward solve---this rectified the non-convergence issues in almost all cases.

%%%%%%%%%%%%%%%%%%
\section{Numerical experiments: FreeGSNKE vs. Fiesta vs. EFIT\texttt{++}} \label{sec:numerics}
\begin{figure}[t!]
    \centering
    \includegraphics[width=0.49\textwidth]{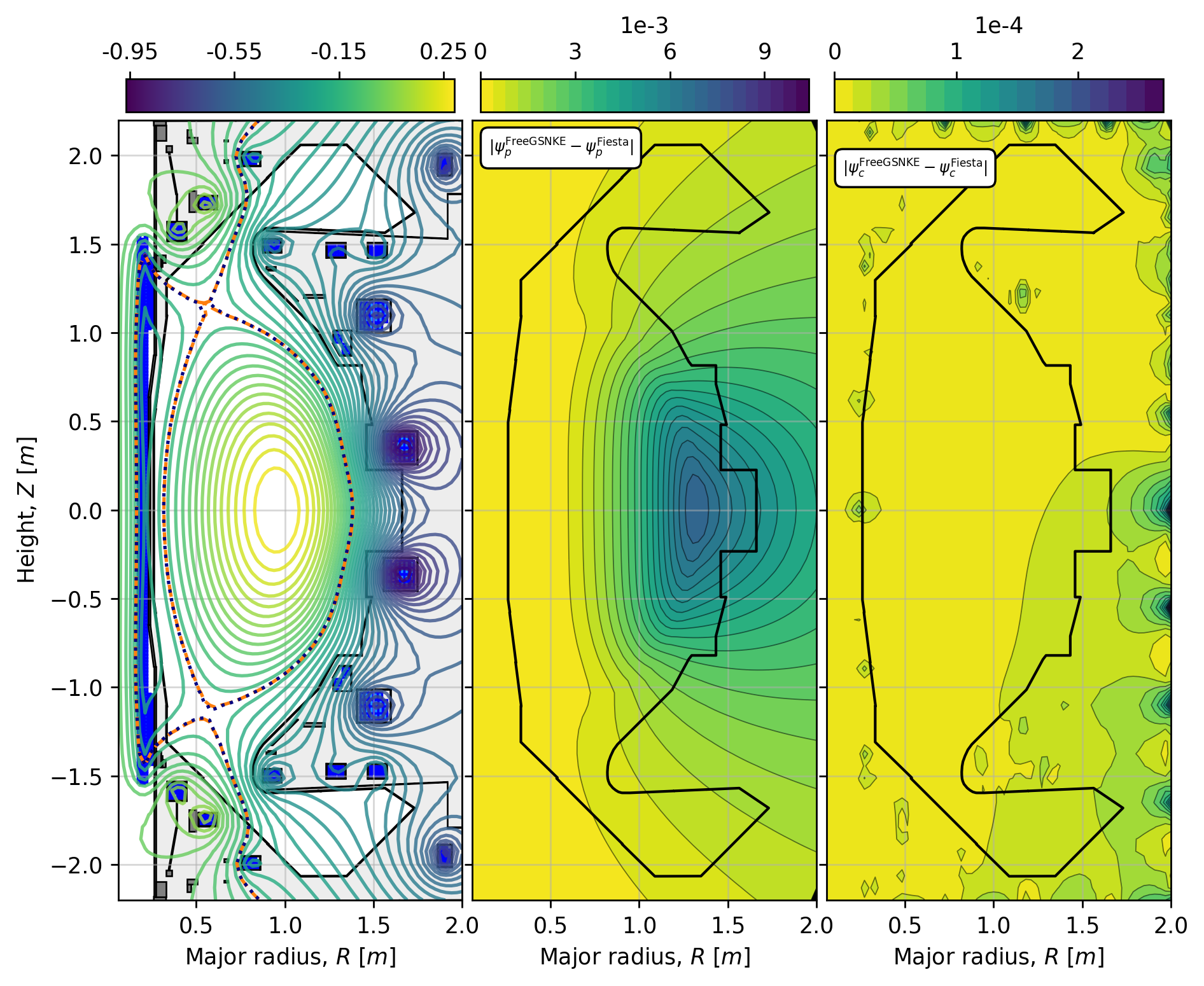}
    \caption{Simulated MAST-U shot 45425 ($t=0.7$s) equilibrium.
    Left: separatrices from both Fiesta (orange) and FreeGSNKE (dotted blue) equilibria alongside the FreeGSNKE $\psi$ contours. 
    Centre: absolute difference in $\psi_{p}$ between the two codes. 
    Right: same as centre but for $\psi_{c}$. 
    }
    \label{fig:exp1_psi}
\end{figure}

% overview
In this section, we compare the equilbria and related shape targets simulated by all three codes across two different MAST-U shots: one with a conventional divertor configuration and one with a Super-X configuration. 
We should reiterate that, although we consider EFIT\texttt{++} to be our reference solver, we have no actual \emph{ground truth} equilibria. 
Therefore, in our comparisons, we measure \emph{differences} between the equilibria produced by the various codes rather than \emph{errors}.
% {\bb All numerical experiments can be found in the public repository:} \url{https://github.com/...}.

% outline of how results were simulated (experimental setup)
We start by briefly outlining the steps taken to obtain these results. 
First, we select the MAST-U shot that we wish to simulate in FreeGSNKE/Fiesta and store\footnote{We should state here that we do not (re-)run EFIT\texttt{++}, we simply extract existing data generated by a post-shot reconstruction stored in the MAST-U database. Data accessed 14/06/24.} the corresponding EFIT\texttt{++} data that we require for each time slice. 
This includes the inputs described in \cref{sec:inputs} and the poloidal flux/shape target output data that we wish to compare to after we have run FreeGSNKE and Fiesta.
After building/loading the machine description in both FreeGSNKE and Fiesta, we then solve the forward GS problem at each time slice sequentially, starting at the first time step for which EFIT\texttt{++} produces a valid GS equilibrium\footnote{During some of the ramp-up and ramp-down of the plasma, EFIT\texttt{++} may struggle to converge to a valid GS equilibrium or produce spurious fits (e.g.\ on the plasma profile coefficients or passive structure currents). 
In these cases, we exclude these time slices from the comparison.}.
For FreeGSNKE, we initialise each simulation with a default initial guess for the plasma flux $\psi_p$---this is obtained automatically by requiring the presence of an O-point in the total flux within the limiter region.
As mentioned before, to ensure convergence, Fiesta is initialised using the $J_p$ field calculated by EFIT\texttt{++}.
While this initialisation is already very close to the desired reference GS solution produced by EFIT\texttt{++}, a comparison with the equilibrium on which Fiesta eventually converges is still informative of the code's performance.
\begin{figure}[t!]
    \centering
    \includegraphics[width=0.49\textwidth]{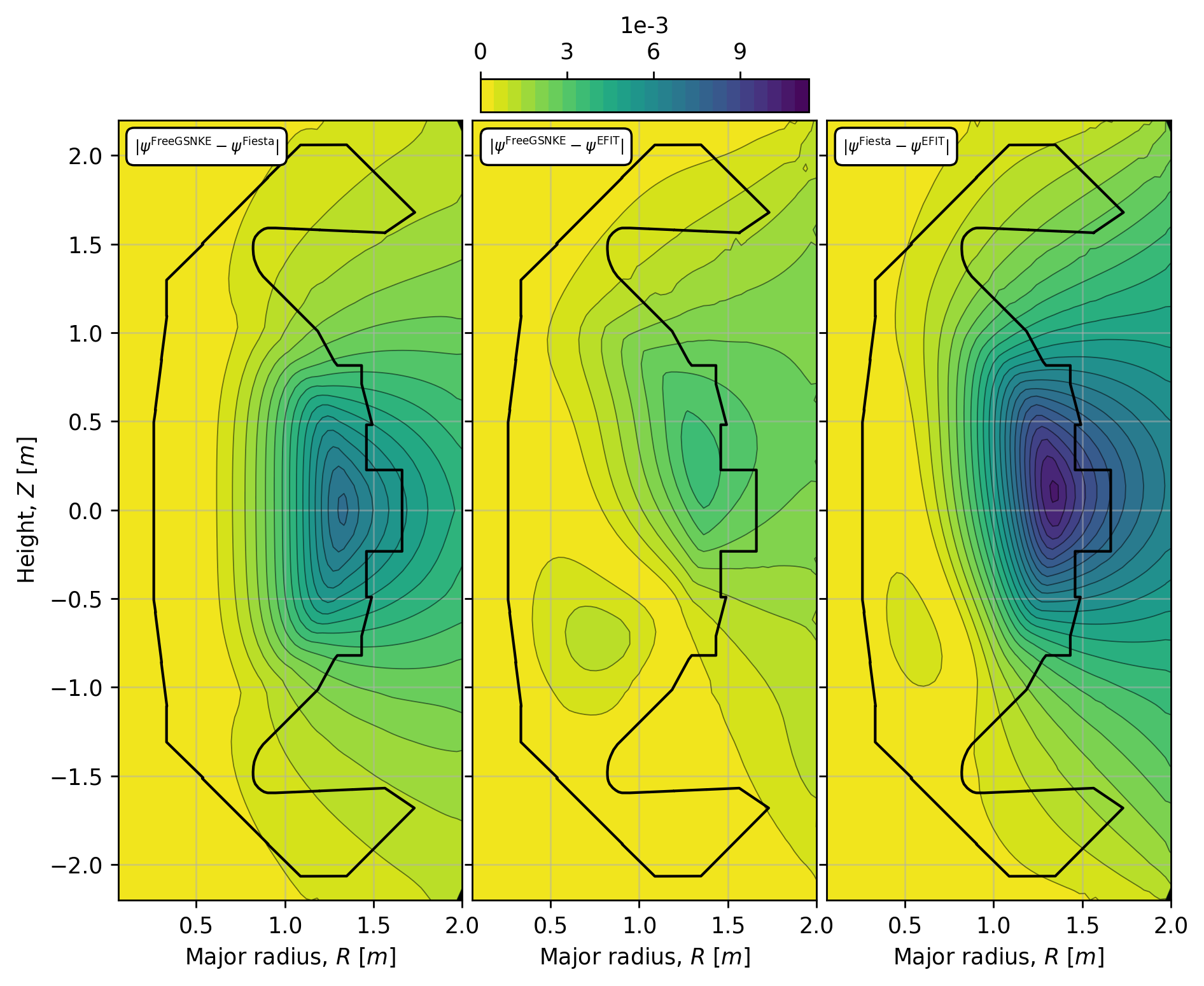}
    \caption{Differences in the total flux $\psi$ between all three codes for MAST-U shot 45425 ($t=0.7$s).
    Left: absolute difference in $\psi$ between FreeGSNKE and Fiesta. 
    Centre: difference between FreeGSNKE and EFIT\texttt{++}.
    Right: difference between Fiesta and EFIT\texttt{++}. 
    }
    \label{fig:exp1_psi_extra}
\end{figure}
\begin{figure*}[t!]
    \centering
    \begin{subfigure}{0.99\linewidth}
        \includegraphics[width=0.96\textwidth]{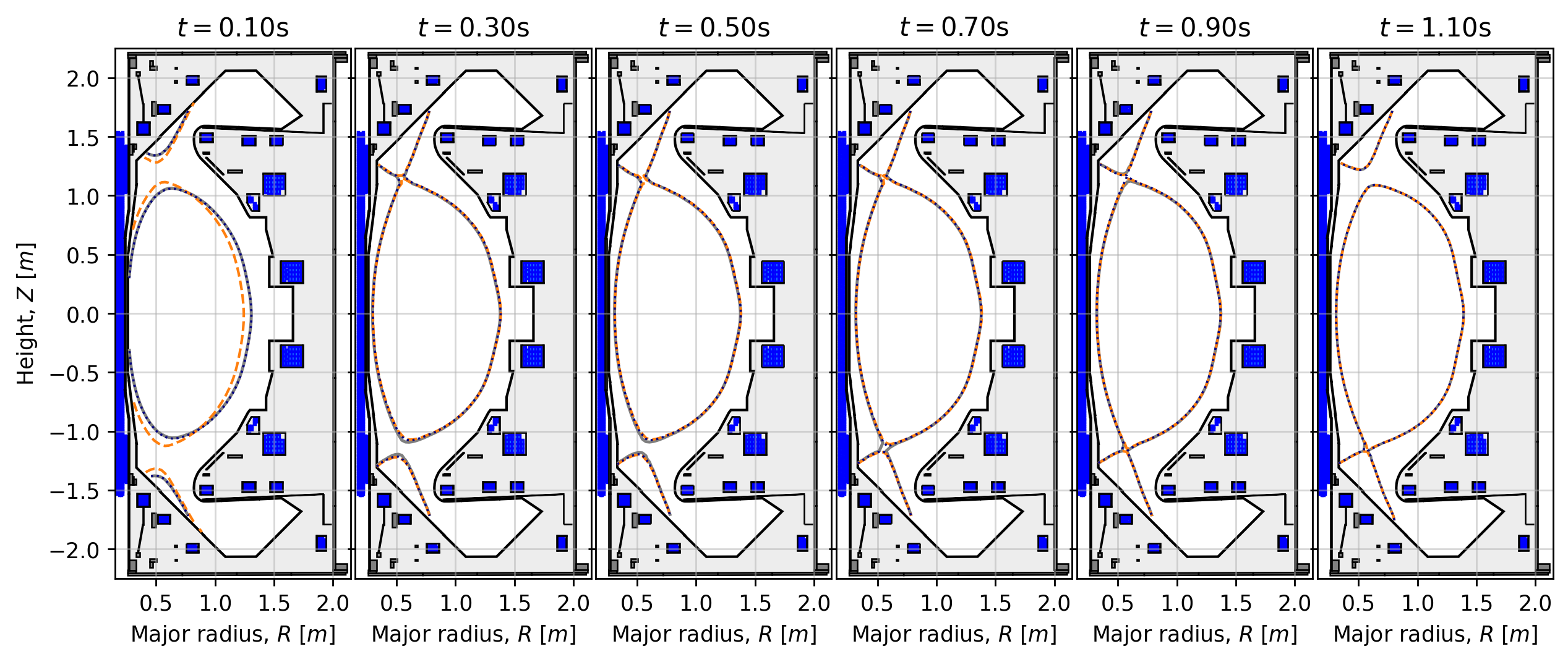}
        % \caption{FreeGSNKE}
    \end{subfigure}
    \begin{subfigure}{0.99\linewidth}
        \includegraphics[width=0.96\textwidth]{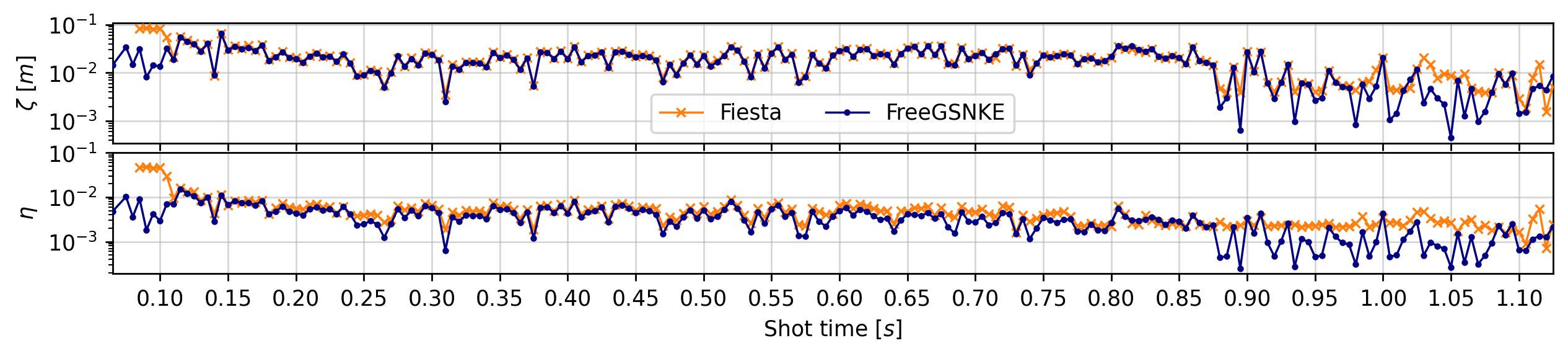}
        % \caption{Fiesta}
    \end{subfigure}
    \caption{Top: evolution of EFIT\texttt{++} (solid grey), Fiesta (dashed orange), and FreeGSNKE (dotted blue) separatrices at different shot times.
    Middle: evolution of the $\zeta$ metric from \eqref{eq:zeta_metric} over time for Fiesta (orange) and FreeGSNKE (blue) compared to EFIT\texttt{++} (divertor legs not included).
    Bottom: similarly, the evolution of the $\eta$ metric from \eqref{eq:eta_metric} over time. 
    }
    \label{fig:exp2_separatrix}
\end{figure*}

%Once the converged value of $\psi_p$ is known, it is can then be used as an initial guess to solve for the next time step, working under the assumption that the plasma flux does not change too drastically between time steps.
%This aids convergence and reduces the number of Newton--Krylov iterations required to solve \eqref{eq:nonlinear_root}. 
% To run Fiesta, we also solve the forward GS problem at each time slice sequentially, except that instead of using $\psi_p$ as an initial guess for the Picard solver, we initialise the iterations using the plasma current density $J_{\phi}$ generated by EFIT\texttt{++}. 

%%%%%%%%%%%%%%%%%%%
\subsection{MAST-U shot 45425: Conventional divertor}

% shot details 
We first simulate MAST-U shot 45425, which has a flat-top plasma current of approximately $750$kA, a double-null shape, and a conventional divertor configuration. 
The plasma is heated using two neutral beam injection systems delivering a total power of approximately $2.5$MW and remains in H-mode confinement for the majority of the shot.

\subsubsection{Single time slice}
Before analysing the entire shot, we wish to briefly discuss and compare a few minor differences between the flux quantities produced by each code for a single time slice of the shot ($t = 0.7$s). 
We begin by comparing FreeGSNKE and Fiesta without EFIT\texttt{++} in \cref{fig:exp1_psi}. 
The left panel displays the $\psi$ contours from FreeGSNKE and an almost perfect overlap of the separatrices from both codes.

\begin{figure*}[t!]
    \centering
    \begin{subfigure}{0.96\linewidth}
        \includegraphics[width=\textwidth]{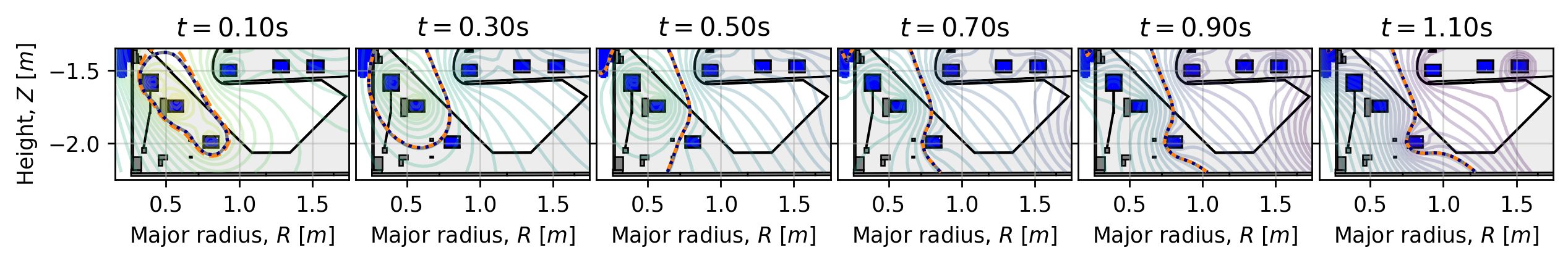}
        % \caption{FreeGSNKE}
    \end{subfigure}
    \begin{subfigure}{0.96\linewidth}
        \includegraphics[width=\textwidth]{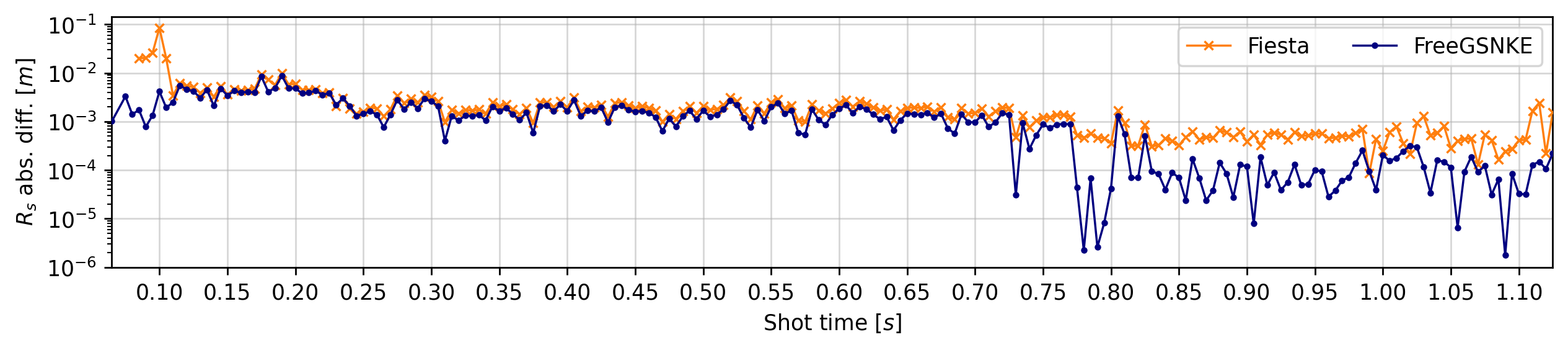}
        % \caption{Fiesta}
    \end{subfigure}
    \caption{Top: evolution of EFIT\texttt{++} (solid grey), Fiesta (dashed orange), and FreeGSNKE (dotted blue) lower divertor strikepoints at different shot times.
    Bottom: absolute difference between between EFIT\texttt{++} and Fiesta/FreeGSNKE for $R_s$ (errors are at the same level for $Z_s$). 
    % Bottom: same as centre but for $z_s$.
    }
    \label{fig:exp2_strikes}
\end{figure*}

We break this down in the centre and right panels. 
The right panel shows the magnitude of differences in $\psi_{c}$, the plasma flux generated by active coils and passive structures. It can be seen that differences are generally small, at a level of $\lesssim 2\negE4$, compared to a total flux that spans a range $\max(\psi) - \min(\psi) \sim 1$. 
These differences appear co-localised with the vessel's passive structures, suggesting they are driven by implementation details in the methods used by either code to distribute the passive structure currents over their poloidal sections\footnote{We also explicitly checked any differences in the flux contribution from the active coils alone and found them to be $\mathcal{O}(10^{-15})$.}.
The central panel shows differences in the plasma flux $\psi_{p}$. The largest differences are localised in the top-right and bottom-right edge pixels of the discretised domain. 
We attribute this to implementation details in the way Fiesta imposes the boundary conditions (see also the right panel of \cref{fig:exp1_psi_extra}, where the same discrepancy is visible again). 
The remaining differences are at a level of $\lesssim 5\negE3$. 
We believe these are largely due to implementation differences in the routines that identify the last closed flux surface of the plasma, rather than being due to the nonlinear solvers themselves. 
In fact, we find that a comparison between the plasma current density distributions $J_p$ calculated by FreeGSNKE and Fiesta for the same total flux $\psi$ results in differences of the same order of magnitude.

% we can see that the absolute differences in the plasma flux $\psi_{p}$ (central plot) are at most $\mathcal{O}(10^{-3})$, where the largest differences are clustered close to the plasma outer radial edge and, looking closely, in the upper/lower right-hand corners of the grid. 
% {\bb In the plasma outer edge region, we suspect that differences in $\psi_p$ are driven by ... differences in calculation of $J_{\phi}$? the boundary condition? I'm not sure ...}
% {\bb In the right-hand side corner grid cells, we are unsure as to what is driving the difference but if we look at \cref{fig:exp1_psi_extra}, we can see that the difference is common to the Fiesta simulation. I'm not sure what the difference is caused by ... possibly by the boundary condition calculation? Or perhaps the spatial discretisation process? Fiesta discretises the elliptic operator using second order-accurate finite differences with a forward sine transform. }

% As we know, the coil flux $\psi_c$ is made up of contributions from both the active coils and the passive structures. The difference in the contribution from the active coils is $\mathcal{O}(10^{-15})$ (not shown) and so what we observe in \cref{fig:exp1_psi} (right) are the differences in the passive structure contributions only. These minor differences are clustered around some of the passive structure locations and are driven mainly by small differences in how the passives are refined in both FreeGSNKE and Fiesta. 

In \cref{fig:exp1_psi_extra}, we compare differences in the total flux $\psi$ between all three codes\footnote{We should note that EFIT\texttt{++} did not produce a breakdown of $\psi$ into $\psi_p$ and $\psi_c$, making the comparison of plasma and conductor flux contributions more difficult.}.
The left panel is similar to the one seen in \cref{fig:exp1_psi} (centre), with differences between FreeGSNKE and Fiesta dominated by differences in $\psi_p$ as just discussed. 
Similarly to the differences between FreeGSNKE and Fiesta, the differences between FreeGSNKE/Fiesta and EFIT\texttt{++} (shown in the centre/right panels, respectively) are largest close to the plasma outer edge, and qualitatively analogous (if not slightly different in magnitude).
\begin{figure}[b!]
    \centering
    \includegraphics[width=0.49\textwidth]{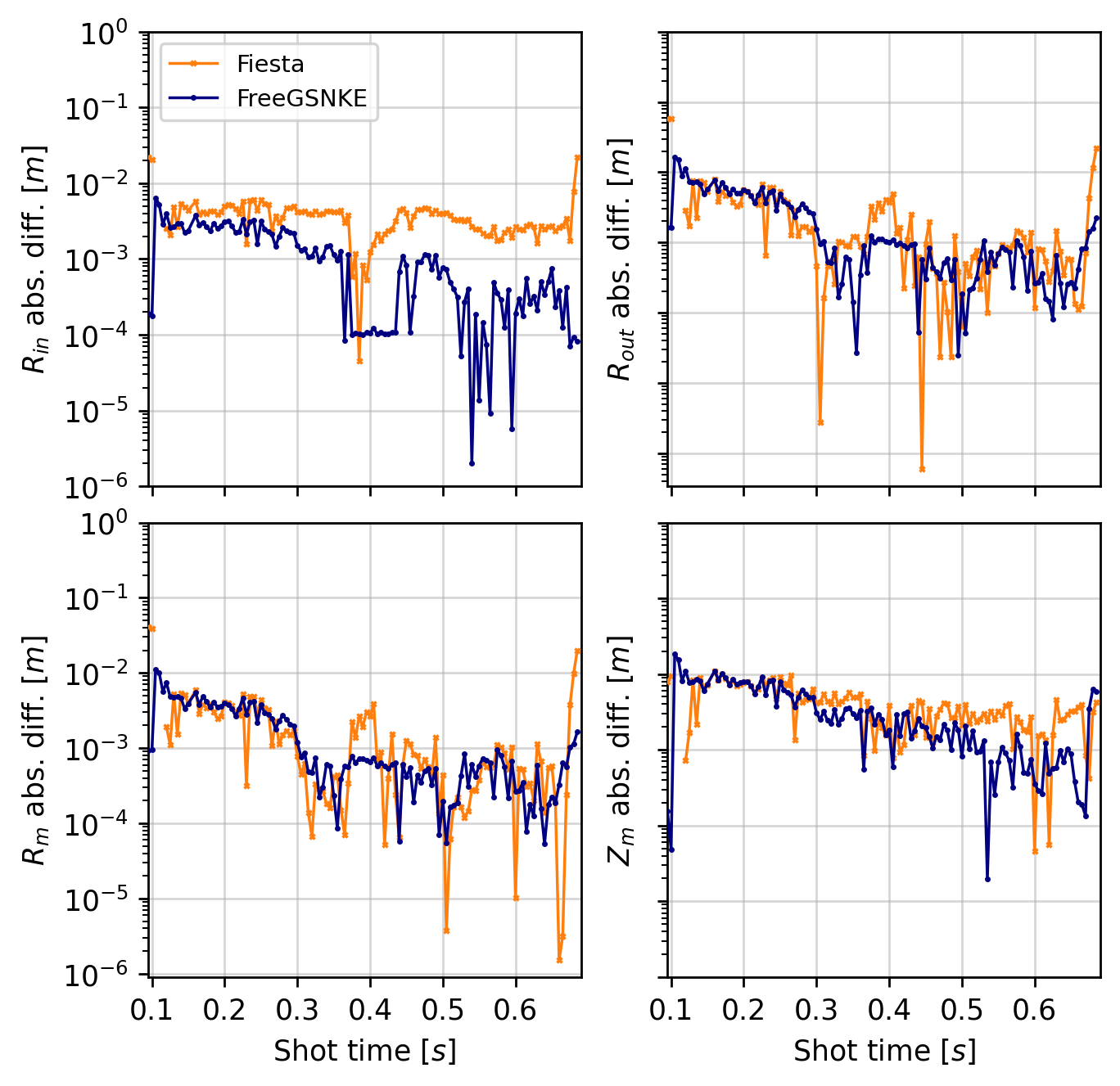}
    \caption{Absolute differences between EFIT\texttt{++} and Fiesta (orange crosses)/FreeGSNKE (blue dots) shape targets. 
    Top: absolute difference in midplane inner $R_{in}$ and outer $R_{out}$ radii. 
    Bottom: absolute difference in magnetic axis components $R_m$ and $Z_m$.
    }
    \label{fig:exp2_shapes}
\end{figure}

% \begin{figure*}[h!]
%     \centering
%     \includegraphics[width=0.99\textwidth]{Figures/exp2_psi.png}
%     \caption{Absolute difference between EFIT\texttt{++} and Fiesta (orange crosses)/FreeGSNKE (blue dots) $\psi$ quantities. 
%     Top: maximum (solid) and median (dashed) absolute difference in $\psi$. 
%     Centre: absolute difference in $\psi_a$. 
%     Bottom: absolute difference in $\psi_b$. 
%     }
%     \label{fig:exp2_psi}
% \end{figure*}

It is worth highlighting explicitly that the mismatches shown in \cref{fig:exp1_psi} and \cref{fig:exp1_psi_extra} are, nominally, beyond the relative tolerance used in both FreeGSNKE and Fiesta---recall this was $\varepsilon = 1\negE6$.
However, as already mentioned: i) differences in the implementation of the passive structures and ii) differences in routines that build the plasma core mask between the three codes at hand, are responsible for introducing mismatches with similar orders of magnitude to those we are seeing. 
Besides, as hinted by the left panel in \cref{fig:exp1_psi} and as we show in the following results, this level of difference has a negligible impact on the shape control targets (and therefore for most practical modelling purposes).
% {\rr This may be a bit too strong, I know I wrote this, just making a note we might want to polish.}

\subsubsection{Entire shot}
% discuss results over whole shot
Over the course of the entire shot, the differences in $\psi$ (for both codes) remain at the levels seen in \cref{fig:exp1_psi_extra}. 
For FreeGSNKE, the median differences in $\psi_a$ and $\psi_b$ over the shot are $1\negE3$ and $6\negE4$, respectively.
For Fiesta, the corresponding differences are on average $3\negE3$ and $1\negE3$, respectively.

In the top panel of \cref{fig:exp2_separatrix}, we plot the evolution of the separatrices produced by all three codes over the lifetime of the shot, observing an excellent match in the majority of shot times.
We plot two metrics in the lower panels of \cref{fig:exp2_separatrix} to compare the core plasma boundary (divertor legs not included).
For both metrics, we find $360$ $(R,Z)$ points that lie on the boundary $\partial \Omega_p$ of each code, where each point is equally spaced in the geometric poloidal angle, centred on the magnetic axis (of the EFIT\texttt{++} core).

The first metric, $\zeta$, quantifies the largest Euclidean distance between corresponding $(R,Z)$ points on the FreeGSNKE/Fiesta boundary and the EFIT\texttt{++} boundary.
This is defined as
\begin{align} \label{eq:zeta_metric}
    \zeta \big(\partial \Omega_{p}^{1}, \partial \Omega_{p}^{2} \big) \defeq \max_{j=1,\dots,360} \big\| \partial \Omega_{p, j}^{1} - \partial \Omega_{p, j}^{2} \big\|_2,
\end{align}
where $\partial \Omega_{p, j}^{1}$ denotes the $j$th $(R,Z)$ point on the boundary from the first code (same for the second code).
An illustration of how this metric is similarly calculated can be found in \citet{stewart2023}[Fig. 2].
We find $\zeta$ to be below $3.5$cm in $95\%$ of cases for FreeGSNKE equilibria (in $96\%$ of cases the median distance is below $1$cm). 
For the case of Fiesta, $\zeta$ is below $4.2$cm for $95\%$ of the time slices (and the median is below $1.13$cm for $96\%$ of cases).

The second metric, $\eta$, first presented by \cite{bardsley2024}, is defined as
\begin{align} \label{eq:eta_metric}
    \eta \big( \Omega_{p}^{1}, \Omega_{p}^{2} \big) \defeq \frac{\big| \Omega_p^{1} \cup  \Omega_p^{2} \big| - \big| \Omega_p^{1} \cap \Omega_p^{2} \big|}{ \big| \Omega_p^{1} \big| + \big| \Omega_p^{2} \big|} \in [0,1],
\end{align}
where $| \cdot |$ denotes the cross-sectional area of a domain in the poloidal plane. 
This dimensionless parameter quantifies the ratio of the total non-overlapping area of the two plasma domains to the sum of their areas.
Values closer to zero indicate a good match between the plasma domains and we can see from the results that both codes yield $\eta \leq 0.01$ for the entire shot (ignoring the early Fiesta time slices).

Any gaps in the time series (and in later plots) are where Fiesta failed to converge to an equilibrium given the tolerance $\varepsilon$---these time slices have been excluded when calculating the quantiles mentioned above.
We would note that during these times slices (where the equilibrium is in a limiter-type configuration), Fiesta can converge when using the \texttt{feedback2} object mentioned in \cref{sec:inputs}.

\begin{figure}[t!]
    \centering
    \includegraphics[width=0.43\textwidth]{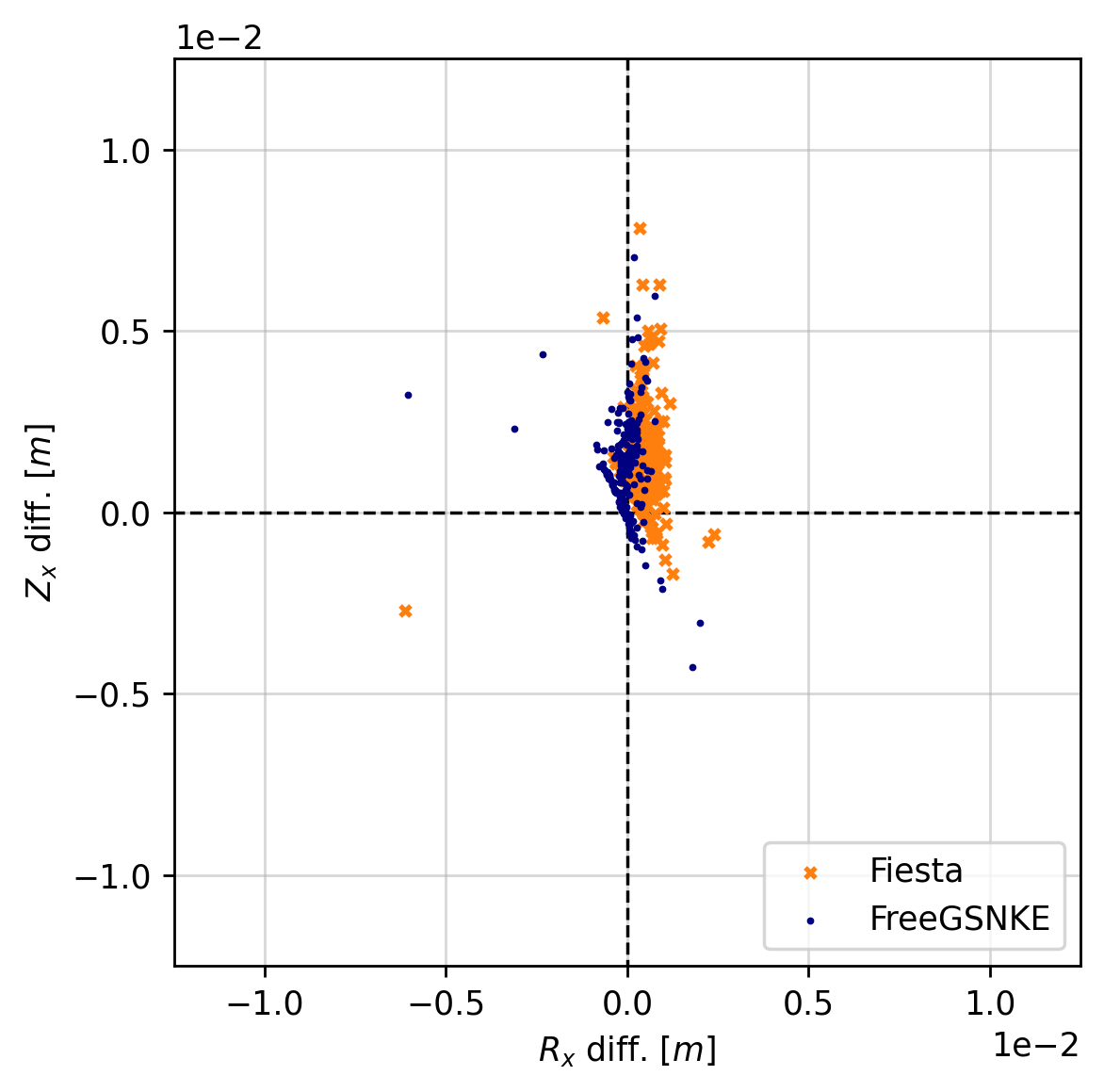}
    \caption{Difference between EFIT\texttt{++} and Fiesta (orange crosses)/FreeGSNKE (blue dots) lower core chamber X-points (over the entire shot). 
    }
    \label{fig:exp2_xpoints}
\end{figure}
\begin{figure}[h!]
    \centering
    \includegraphics[width=0.45\textwidth]{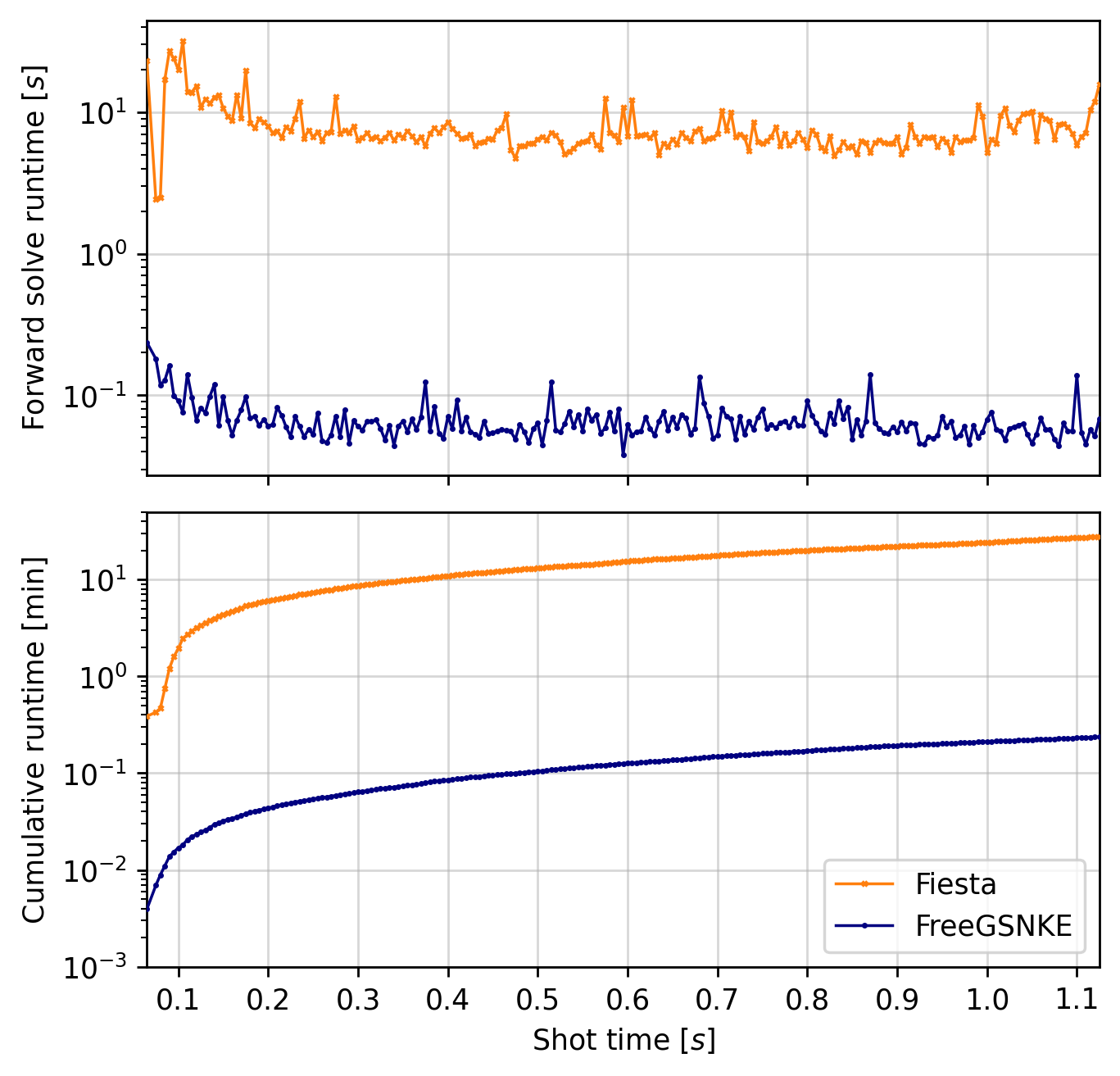}
    \caption{Fiesta (orange) and FreeGSNKE (blue) forward solve runtimes over the shot (in seconds). 
    Top: runtime per forward solve. 
    Bottom: cumulative runtime of forward solves (in total, Fiesta takes \SI{27}{\minute} \SI{48}{\second} and FreeGSNKE takes \SI{16}{\second}.
    }
    \label{fig:exp2_runtimes}
\end{figure}
\begin{figure*}[t!]
    \centering
    \begin{subfigure}{0.95\linewidth}
        \includegraphics[width=\textwidth]{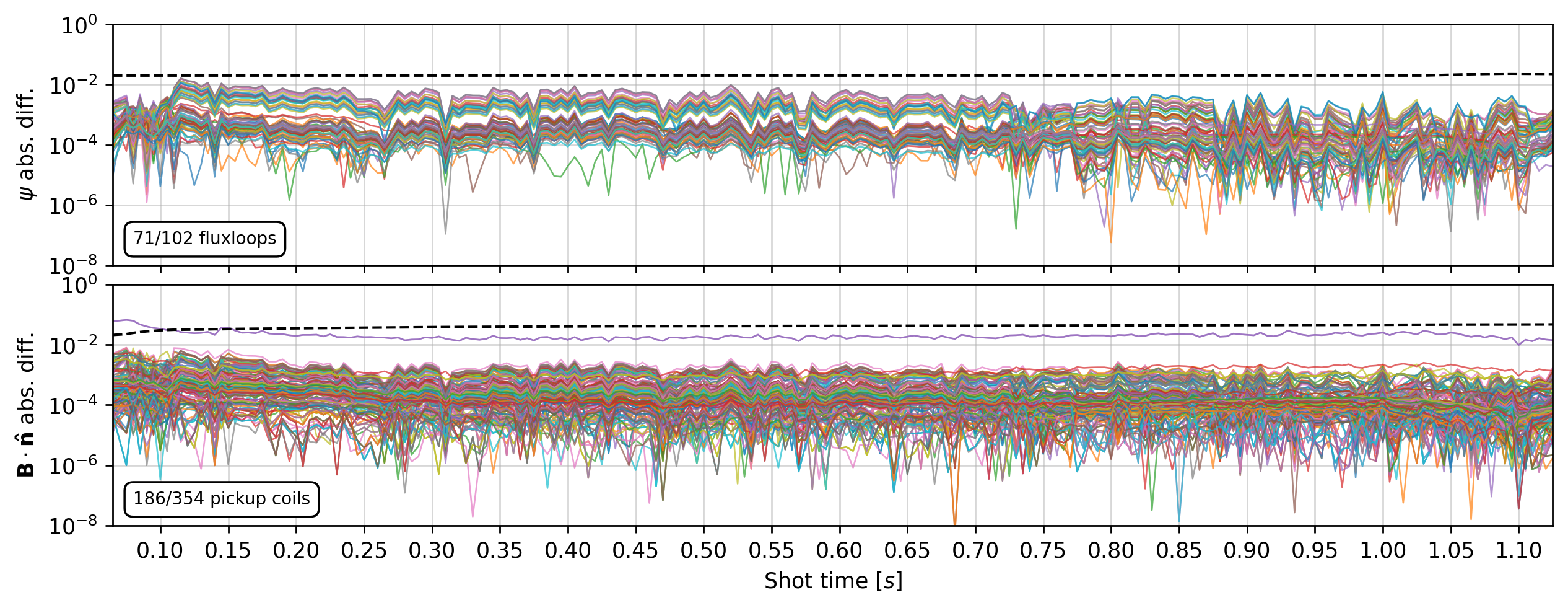}
        % \caption{FreeGSNKE}
    \end{subfigure}
    \caption{Absolute differences between FreeGSNKE and EFIT\texttt{++} magnetics diagnostics. 
    Top: absolute difference in poloidal flux $\psi$, measured at $71$ out of a total $102$ fluxloops. 
    Bottom: absolute difference in magnetic field strength $\bm{B} \cdot \bm{\hat{n}}$, measured at $186$ out of a total $354$ pickup coils.
    Each multicoloured line represents an individual diagnostic while the dashed black line represents the (maximum) standard deviation associated with any diagnostic in EFIT\texttt{++} at a given time slice (i.e. the acceptable level of difference we should find between FreeGSNKE and EFIT\texttt{++}).
    }
    \label{fig:exp2_magnetics}
\end{figure*}

\begin{figure*}[h!]
    \centering
    \begin{subfigure}{0.96\linewidth}
        \includegraphics[width=\textwidth]{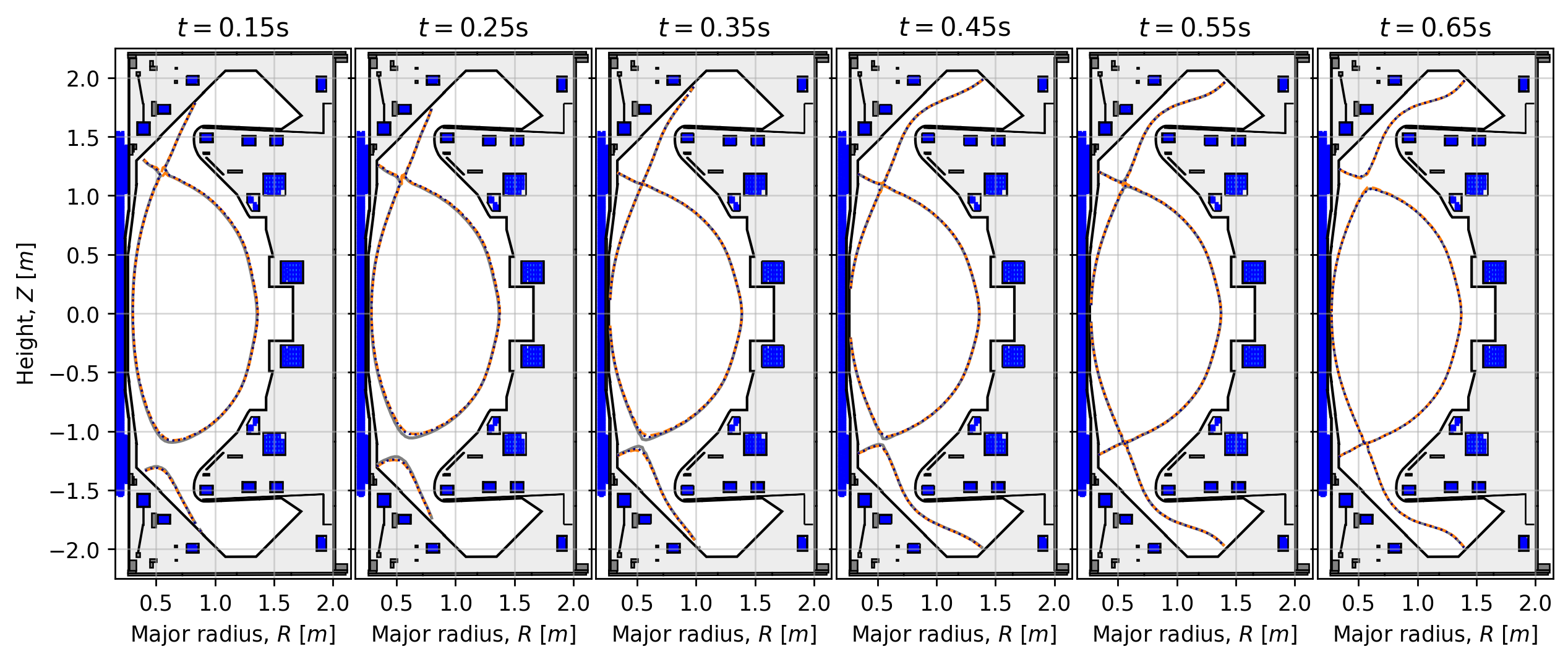}
        % \caption{FreeGSNKE}
    \end{subfigure}
    \begin{subfigure}{0.96\linewidth}
        \includegraphics[width=\textwidth]{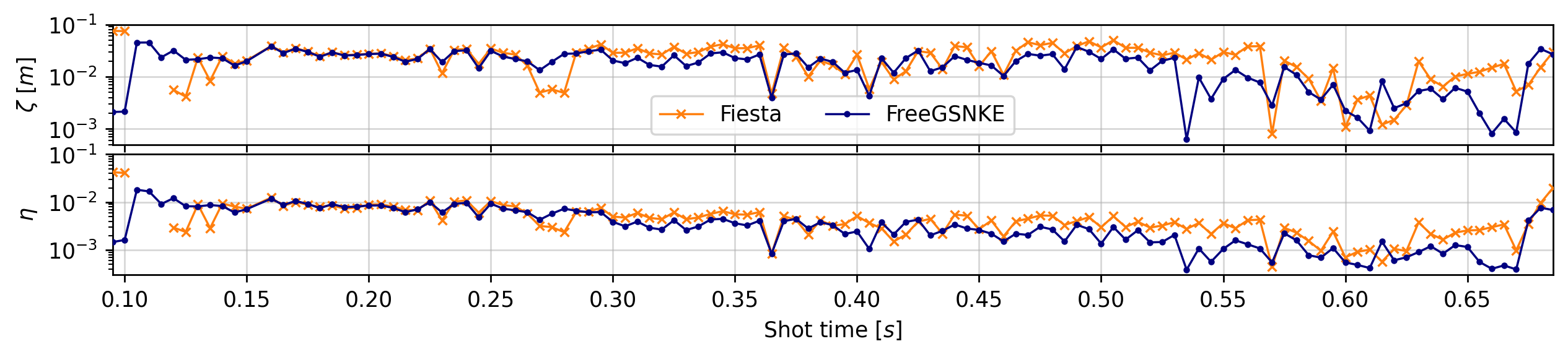}
        % \caption{Fiesta}
    \end{subfigure}
    \caption{Top: evolution of EFIT\texttt{++} (solid grey), Fiesta (dashed orange), and FreeGSNKE (dotted blue) separatrices at different shot times.
    Middle: evolution of the $\zeta$ metric from \eqref{eq:zeta_metric} over time for Fiesta (orange) and FreeGSNKE (blue) compared to EFIT\texttt{++} (divertor legs not included).
    Bottom: similarly, the evolution of the $\eta$ metric from \eqref{eq:eta_metric} over time. 
    }
    \label{fig:exp3_separatrix}
\end{figure*}

\begin{figure*}[t!]
    \centering
    \begin{subfigure}{0.95\linewidth}
        \includegraphics[width=\textwidth]{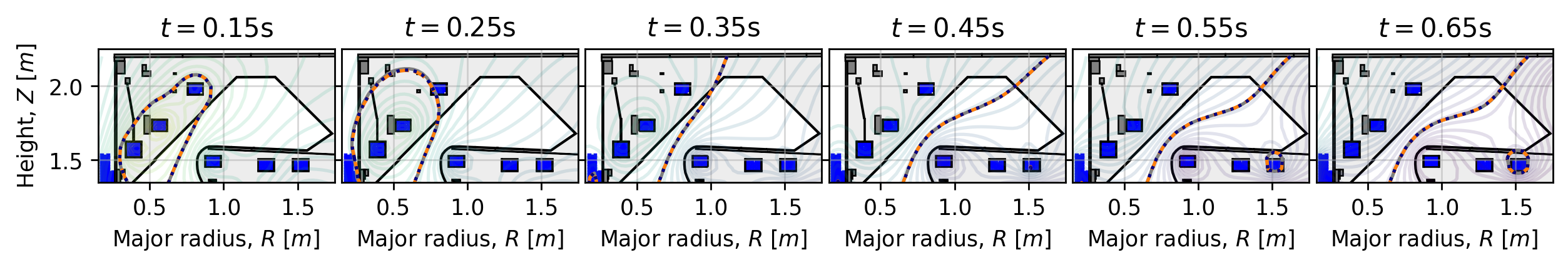}
        % \caption{FreeGSNKE}
    \end{subfigure}
    \begin{subfigure}{0.95\linewidth}
        \includegraphics[width=\textwidth]{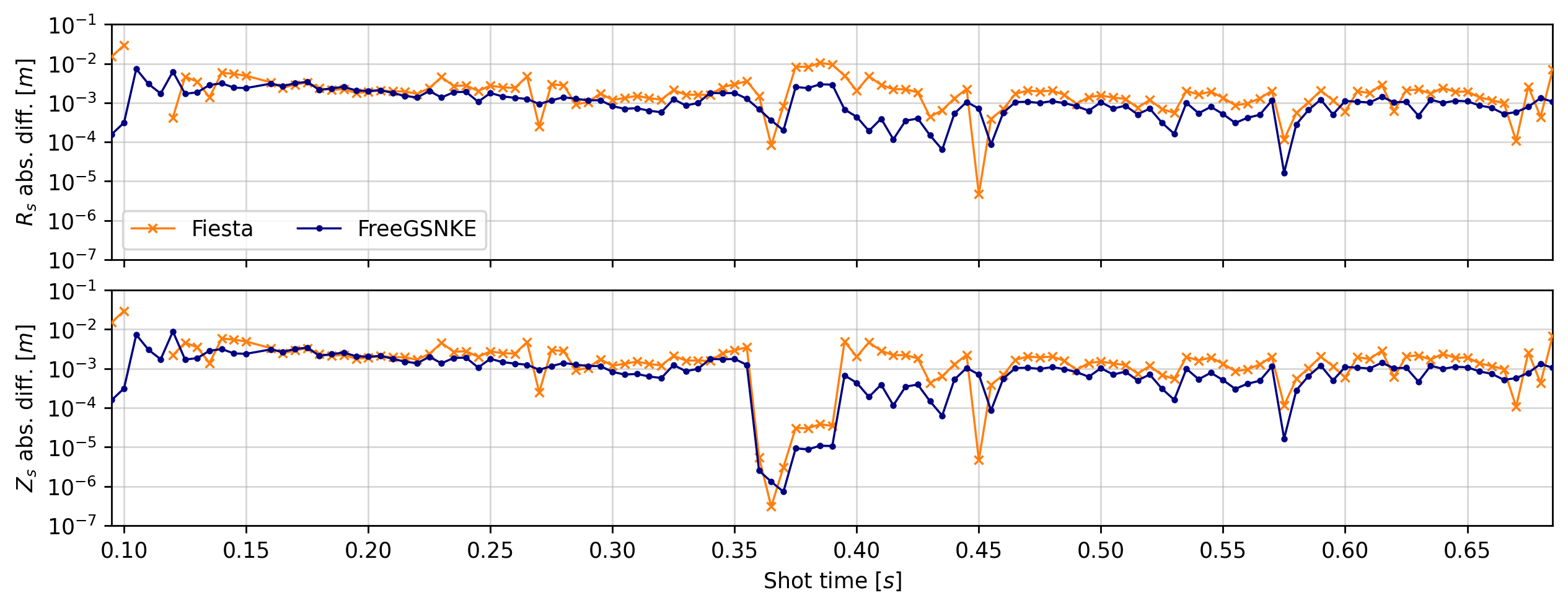}
        % \caption{Fiesta}
    \end{subfigure}
    \caption{Top: evolution of EFIT\texttt{++} (solid grey), Fiesta (dashed orange), and FreeGSNKE (dotted blue) lower divertor strikepoints at different shot times.
    Centre: absolute difference between between EFIT\texttt{++} and Fiesta/FreeGSNKE for $R_s$. 
    Bottom: same as centre but for $Z_s$.
    }
    \label{fig:exp3_strikes}
\end{figure*}
In \cref{fig:exp2_strikes} we monitor the evolution of a strikepoint along the lower divertor tiles, again, observing an excellent match from both codes. 
The difference in $(R_s,Z_s)$ compared to EFIT\texttt{++} is less than a centimetre for most of the shot---very early times being the exception in the case of Fiesta. 
Note that while the difference in $Z_s$ is not explicitly shown, it is of the same order as that of $R_s$. 

In \cref{fig:exp2_shapes}, we plot the differences in the magnetic axis $(R_m, Z_m)$, the midplane inner radius $R_{in}$, and the midplane outer radius $R_{out}$ (recall \cref{fig:MASTU}). 
Typical differences in the inner/outer radii are of the order of millimetres for FreeGSNKE, with only marginally higher values for Fiesta.
With respect to the magnetic axis, differences in both codes vs. EFIT\texttt{++} track one another to sub-centimetre precision as well.  
There are, however, some isolated times during the initial phase of the ramp up where the differences between Fiesta and EFIT\texttt{++} are significantly higher (see similar differences in later plots) than in later time slices.
While FreeGSNKE has found (precisely) the EFIT\texttt{++} GS equilibria during these early slices, we suspect that Fiesta may have simply found another set of (equally valid) GS equilibria.
The physical difference in these (limited, not yet diverted) equilibria can be seen more clearly in \cref{fig:exp2_separatrix} at $t=0.10s$.
While we do not have a conclusive reason for the presence of multiple GS equilibria here (given the same input parameters), identifying under what conditions these equilibria co-exist may be worth further investigation \citep{ham2024}.

In \cref{fig:exp2_xpoints}, we plot the difference in the lower core chamber X-point over the shot. 
As we did for the separatrix calculations, we identify all X-points for each code's equilibria using FreeGSNKE's built-in functionality (calculated by finding saddle points of $\psi)$. 
This was because Fiesta and EFIT\texttt{++} would inconsistently return only a single X-point, sometimes in the lower core chamber, sometime the upper.
There appears to be a small vertical bias towards higher X-points by a few millimetres in both Fiesta and FreeGSNKE---also visible in some of the upper panels of \cref{fig:exp2_separatrix}.
Despite this slight bias, both codes are accurate with respect to EFIT\texttt{++} to within half a centimetre for $98\%$ and $97\%$ of the shot, respectively ($100\%$ are within $1$cm). 

The runtimes, both per time slice and cumulatively across the shot, are displayed in \cref{fig:exp2_runtimes}.
Median runtimes are \SI{6.7}{\second} per forward solve for Fiesta, \SI{0.07}{\second} for FreeGSNKE.
Fiesta is significantly hindered by the use of the \texttt{feedback3} object which demands a second nonlinear solver loop to stabilise the Picard iterations\footnote{However, if one is willing to allow the P6 coil current(s) be modified, Fiesta can run much faster using the \texttt{feedback2} object instead---the median runtime was around \SI{0.22}{\second} per forward solve in this case.}.
FreeGSNKE makes use of the faster and more stable convergence of the Newton--Krylov method, as well as using \texttt{numba} just-in-time compilation for some core routines.
% {\bb and what else makes it fast? Just-in-time compilation?}
% We should, however, recall from comments in \cref{sec:inputs} that Fiesta is significantly hindered by the use of the \texttt{feedback3} object which demands a second nonlinear solver loop to stabilise the Picard iterations. 
As mentioned before, all equilibria were simulated sequentially with both codes. 
However, given the independence of each time slice, nothing prevents an embarrassingly parallel implementation---though this was not an objective in this paper. 

Finally in \cref{fig:exp2_magnetics}, we compare readings from the fluxloops and pickup coils in FreeGSNKE with those used by EFIT\texttt{++} to reconstruct the equilibria.
Note that the measurements are only compared on a fraction of the total number of fluxloops and pickup coils in MAST-U because EFIT\texttt{++} automatically excludes diagnostics which have failed or whose readings exceed a certain calibration threshold \citep{kogan2022, ryan2023}.
As expected, we can see that the absolute difference between readings from both sets of diagnostics is very small and below the (maximum) level of standard deviation assigned to each diagnostic during the EFIT\texttt{++} reconstruction (i.e. below the dashed black line).
The one exception to this was one of the pickup coils (located between the upper P6 and DP coils) that can be seen exceeding this threshold during the ramp-up and remaining consistently higher than other pickups throughout the shot.
The measurements from this pickup coil could have been affected by calibration issues or perhaps maybe evaded the EFIT\texttt{++} faulty probe detection.

% \begin{figure*}[h!]
%     \centering
%     \begin{subfigure}{0.95\linewidth}
%         \includegraphics[width=\textwidth]{Figures/exp3_magnetics.png}
%         % \caption{FreeGSNKE}
%     \end{subfigure}
%     \caption{{\bb Absolute differences between FreeGSNKE and EFIT\texttt{++} magnetics diagnostics. 
%     Top: absolute difference in poloidal flux $\psi$, measured at $70$ out of a total $102$ fluxloops. 
%     Bottom: absolute difference in magnetic field strength $\bm{B} \cdot \bm{\hat{n}}$, measured at $186$ out of a total $354$ pickup coils.
%     Each multicoloured line represents an individual diagnostic while the dashed black line represents the (maximum) standard deviation associated with any diagnostic in EFIT\texttt{++} at a given time slice (i.e. the acceptable level of difference we should find between FreeGSNKE and EFIT\texttt{++}).
%     }
%     }
%     \label{fig:exp3_magnetics}
% \end{figure*}

% \begin{figure*}[h!]
%     \centering
%     \includegraphics[width=0.99\textwidth]{Figures/exp3_psi.png}
%     \caption{Absolute differences between EFIT\texttt{++} and Fiesta (orange crosses)/FreeGSNKE (blue dots) $\psi$ quantities. 
%     Top: maximum (solid) and median (dashed) absolute difference in $\psi$. 
%     Centre: absolute difference in $\psi_a$. 
%     Bottom: absolute difference in $\psi_b$. 
%     }
%     \label{fig:exp3_psi}
% \end{figure*}

\subsection{MAST-U shot 45292: Super-X divertor}

% shot details 
Here, we focus on MAST-U shot $45292$, which has a flat-top plasma current of approximately $750$kA, a double-null shape, and a Super-X divertor configuration. 
The plasma is Ohmically heated, i.e.\ there is no neutral beam heating, and remains in the L-mode confinement regime throughout the shot. 
\begin{figure}[b!]
    \centering
    \includegraphics[width=0.49\textwidth]{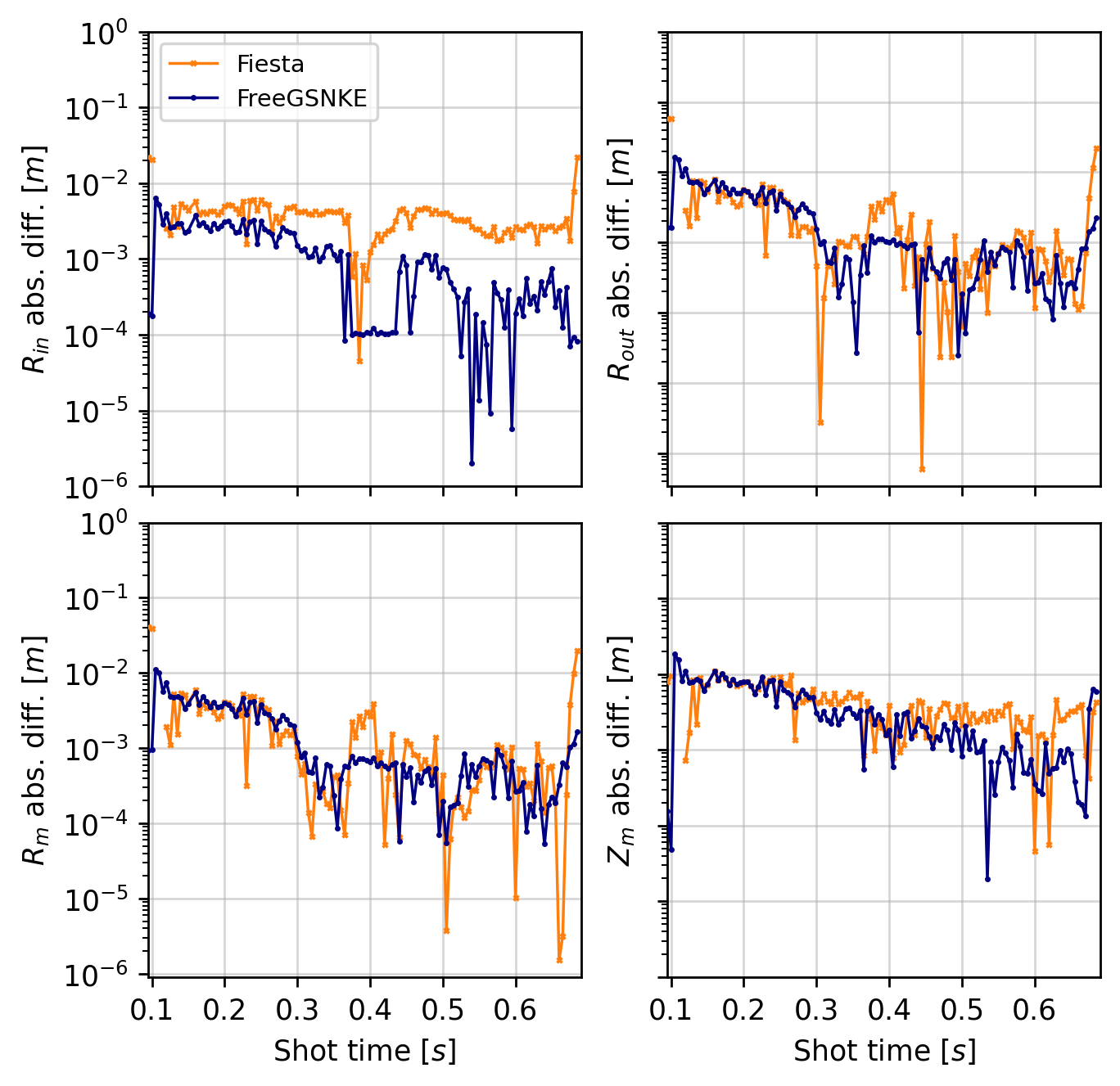}
    \caption{Absolute differences between EFIT\texttt{++} and Fiesta (orange crosses)/FreeGSNKE (blue dots) shape targets. 
    Top: absolute difference in midplane inner $R_{in}$ and outer $R_{out}$ radii. 
    Bottom: absolute difference in magnetic axis components $R_m$ and $Z_m$. 
    }
    \label{fig:exp3_shapes}
\end{figure}
\begin{figure}[t!]
    \centering
    \includegraphics[width=0.43\textwidth]{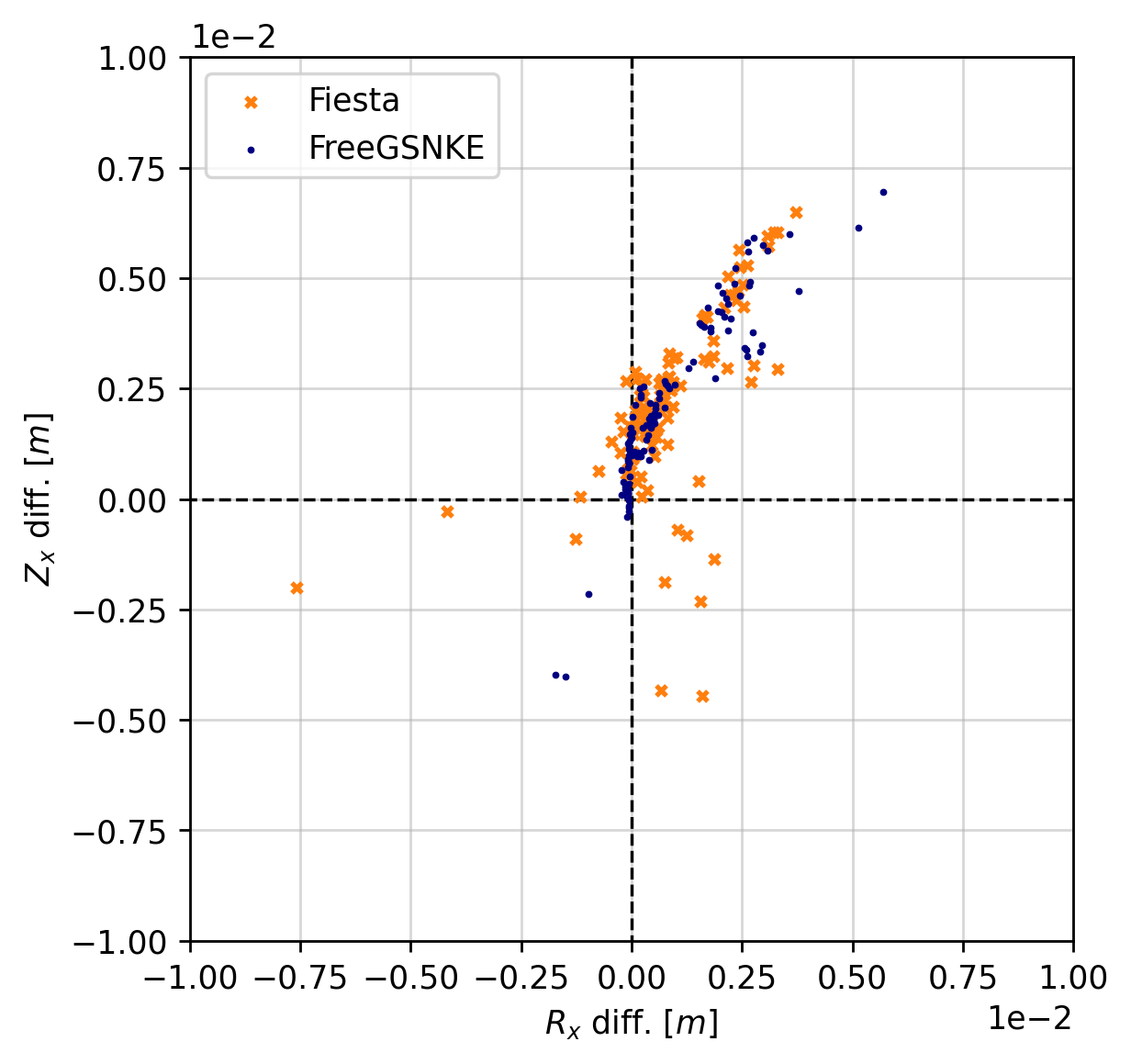}
    \caption{Absolute differences between EFIT\texttt{++} and Fiesta (orange crosses)/FreeGSNKE (blue dots) lower core chamber X-points (over entire shot).
    }
    \label{fig:exp3_xpoints}
\end{figure}

% results discussion (what's different from the previous shot?)
As we did for the previous shot, we plot the evolution of the separatrices from all three codes over time in \cref{fig:exp3_separatrix}, again discerning a very good agreement between all time slices (including qualitatively on the divertor legs). 
Again, $\zeta$ reveals centimetre level differences in the core boundaries (in the worst cases) and $\eta$ returns values approximately less than $0.01$.
Given the Super-X configuration, the upper divertor strikepoint now evolves across the tiles.
We can see, in \cref{fig:exp3_strikes}, good agreement with differences in $R_s$ and $Z_s$ remaining at similar levels (though marginally different).

Differences in poloidal fluxes remain at similar levels as seen in the conventional divertor shot while shape targets again match to sub-centimetre precision---see \cref{fig:exp3_shapes}.
Upper core chamber X-points from FreeGSNKE and Fiesta are again accurate to within half a centimetre for $92\%$ and $87\%$ of time slices shown ($100\%$ within $1$cm), respectively (see \cref{fig:exp3_xpoints}).
Runtimes for both simulations were almost identical to those seen in the previous experiment and are therefore not shown again. 
To see the additional results not shown here for the Super-X shot, refer to the code repository.

% \begin{figure}[h!]
%     \centering
%     \includegraphics[width=0.45\textwidth]{Figures/exp3_runtimes.png}
%     \caption{{\bb Could remove this plot? Don't need runtimes twice.} Fiesta (orange) and FreeGSNKE (blue) forward solve runtimes over the shot (in seconds). 
%     Top: runtime per forward solve. 
%     Bottom: cumulative runtime of forward solves (in total, Fiesta takes \SI{22}{\minute} \SI{56}{\second} and FreeGSNKE takes \SI{13}{\second}.
%     }
%     \label{fig:exp3_runtimes}
% \end{figure}

% %%%%%%%%%%%%%%%%%%%%%%%%%%%%%%%%%%%%%%%%%%%%%%%%%%%%%%%%%%%%%%%%%
\section{Conclusions} \label{sec:discussion}

% what did we do and what can we conclude?
In this paper, we have demonstrated that the static forward GS solvers (see \cref{sec:forward_problem}) in FreeGSNKE and Fiesta can accurately reproduce equilibria generated by magnetics-only EFIT\texttt{++} reconstructions on MAST-U.

% machine description
To achieve this, we began (\cref{sec:inputs}) by outlining which features of the MAST-U machine would be included in the forward solver machine descriptions, using those that most closely matched the one by EFIT\texttt{++}.
We highlighted the capability of both codes being able to model the active poloidal field coils as either up/down symmetric or asymmetric, noting that EFIT\texttt{++} uses the asymmetric setting. 
In addition, both codes have the option to refine passive structures into smaller filaments in order to distribute the induced current in them across their surface areas for better electromagnetic modelling. 
Following this, we set the conductor currents and prescribed appropriate plasma current density profiles---in this case the polynomial-based ``Lao'' profiles whose coefficients are determined by EFIT\texttt{++}.
In addition to some other code-specific parameters, we then used this computational pipeline to begin simulating the MAST-U equilibria.

% recap and discuss our results (what can we conclude/takeaway)?
The poloidal flux quantities and shape targets generated by the FreeGSNKE and Fiesta simulations in \cref{sec:numerics} show excellent agreement with the corresponding quantities from EFIT\texttt{++} for both MAST-U shots tested. 
More specifically, separatrices from both codes match those of EFIT\texttt{++}, with the largest distances between the core boundaries found to be on the order of centimetres.
Strikepoints, X-points, magnetic axes, and inner/outer midplane radii differences between the codes were simulated to sub-centimetre precision.

% future work
The static GS solver in FreeGSNKE has now been validated against both analytic equilibria (see \cite{amorisco2024}) and experimental reconstructions from MAST-U (this paper). 
It has also been shown to produce numerically equivalent equilbria to the Fiesta code, which itself has been validated on experimentally reconstructed equilibria from a number of different tokamak devices (refer back to \cref{sec:intro}).
Given its user-friendly Python interface and superb computational speed/accuracy, we hope that this work will enable the more widespread adoption of FreeGSNKE for machine learning-based plasma control (e.g.\ building libraries of plasma equilbria) and for power plant design optimisation studies (e.g.\ identifying optimal poloidal field coil or magnetic diagnostic locations).
We also stress that FreeGSNKE itself is not designed (and therefore not intended) for use in real-time plasma control.
Furthermore, we hope that the code and datasets made available with this paper will be of use in validation studies for other equilibrium modelling codes.

Some avenues for future work include validating the dynamic forward GS solver in FreeGSNKE using real-world plasma reconstructions, incorporating more complex/unconstrained plasma profile functions, and making use of data assimilation techniques to carry out probabilistic (uncertainty-aware) equilibrium reconstruction.

% %%%%%%%%%%%%%%%%%%%%%%%%%%%%%%%%%%%%%%%%%%%%%%%%%%%%%%%%%%%%%%%%%
% Acknowledgements

\section*{Acknowledgements}
The authors would like to thank Stephen Dixon and Oliver Bardsley (UKAEA) for some very helpful discussions around the MAST-U data handling and for help in quantifying differences in the plasma boundaries. 
We would also like to thank Ben Dudson (LLNL) for pointing us in the direction of a number of very useful FreeGS references. 
This work was part-funded by the FARSCAPE project, a collaboration between UKAEA and the UKRI-STFC Hartree Centre, and by the EPSRC Energy Programme (grant number EP/W006839/1). To obtain further information, please contact PublicationsManager@ukaea.uk.

For the purpose of open access, the author(s) has applied a Creative Commons Attribution (CC BY) licence to any Author Accepted Manuscript version arising.

% %%%%%%%%%%%%%%%%%%%%%%%%%%%%%%%%%%%%%%%%%%%%%%%%%%%%%%%%%%%%%%%%%
% Data availability

\section*{Data availability}
The code scripts and data used in this paper can be found here: \url{https://doi.org/10.14468/26m5-ey02}.

% %%%%%%%%%%%%%%%%%%%%%%%%%%%%%%%%%%%%%%%%%%%%%%%%%%%%%%%%%%%%%%%%%
% Declarations

\section*{Declarations}
The authors have no conflicts of interest to declare.

%%%%%%%%%%%%%%%%%%%%%%%%%%%%%%%%%%%%%%%%%%%%%%%%%%%%%%%%%%%%%%%%%
% Appendices

\begin{appendices} \label{appendix}
\crefalias{section}{appendix}
%\numberwithin{equation}{section}

\section{Alternate profile functions: tension spline} \label{app:tension_spline}

The pressure and toroidal current profiles for an EFIT\texttt{++} reconstruction that uses both magnetics and MSE measurements are defined by
\begin{align*}
    \frac{\mathrm{d} p}{\mathrm{d} \tilde{\psi}} = \sum_{i=0}^{n_p - 1} f_i(\tilde{\psi}) \quad \text{and} \quad 
    F \frac{\mathrm{d} F}{\mathrm{d} \tilde{\psi}} = \sum_{i=0}^{n_F - 1} f_i(\tilde{\psi})
\end{align*}
where $f_i(\tilde{\psi})$ are the \emph{tension spline} (basis) functions \citep{costantini1999, boulila2021}.
Note that we have slightly abused notation as the functions $f_i$ are not the same for each profile function, they each depend on different parameters (see below).
They are defined as
\begin{align*}
    f_i(\hat{\psi}) = 
    \begin{cases} 
        g_i^{(1)}(\hat{\psi}) + g_i^{(2)}(\hat{\psi}) + g_i^{(3)}(\hat{\psi}) & \hat{\psi} \in [\hat{\psi}_i, \hat{\psi}_{i+1}), \\
        0 & \text{otherwise},
    \end{cases}
\end{align*}
where 
\begin{align*}
    g_i^{(1)}(\hat{\psi}) &= y_i \frac{\hat{\psi}_{i+1} - \hat{\psi}}{\hat{\psi}_{i+1} - \hat{\psi}_i} + y_{i+1} \frac{\hat{\psi} - \hat{\psi}_i}{\hat{\psi}_{i+1} - \hat{\psi}_i}, \\
    g_i^{(2)}(\hat{\psi}) &= \frac{z_i}{\sigma^2} \Bigg[ 
        \frac{\sinh(\sigma (\hat{\psi}_{i+1} - \hat{\psi}) )}{\sinh(\sigma (\hat{\psi}_{i+1} - \hat{\psi}_i) )} - \frac{\hat{\psi}_{i+1} - \hat{\psi}}{\hat{\psi}_{i+1} - \hat{\psi}_i} \Bigg], \\
    g_i^{(3)}(\hat{\psi}) &= \frac{z_{i+1}}{\sigma^2} \Bigg[ \frac{\sinh(\sigma (\hat{\psi} - \hat{\psi}_i) )}{\sinh(\sigma (\hat{\psi}_{i+1} - \hat{\psi}_i) )} - \frac{\hat{\psi} - \hat{\psi}_i}{\hat{\psi}_{i+1} - \hat{\psi}_i} \Bigg].
\end{align*}
Note that the final function $f_{n_p-1}$ is defined over the entire interval $[\hat{\psi}_{n_p-1}, \hat{\psi}_{n_p}]$ (similarly for $f_{n_F-1}$). 

As mentioned, the parameters in each $f_i$ are different for both profile functions (hence the abuse of notation) and are defined as follows:
\begin{itemize}[label=$\circ$] %, labelsep=0.5em, left=1em, itemsep=0.5em]
    \item $\hat{\psi}_i$ are the knot points (i.e.\ locations in interval $[0,1]$). 
    \item $y_i = f(\hat{\psi}_i)$ are the values of the profile function at the knot points. 
    \item $z_i = f''(\hat{\psi}_i)$ are the values of the second derivative of the profile function at the knot points.
    \item $\sigma > 0$ is the tension parameter (sending $\sigma \rightarrow \infty$ will results in a piecewise linear spline, whilst smaller $\sigma$ will results in a smoother spline).
\end{itemize}

Both the knot points and the tension parameter are provided as inputs to EFIT\texttt{++} so that it can calculate (fit) the profile function values and second derivatives at the knot points.
These parameters are all that are required to reconstruct $p'$ and $FF'$.
The tension spline enables one to specify more complex profile function shapes and, as with the polynomial profiles, different boundary (and sometimes internal) conditions can be enforced by providing EFIT\texttt{++} with the relevant constraints on the parameter fits (i.e.\ constraints on specific $y_i$ and $z_i$).

\end{appendices}

%%%%%%%%%%%%%%%%%%%%%%%%%%%%%%%%%%%%%%%%%%%%%%%%%%%%%%%%%%%%%%%%%

% Bibliography

\begingroup
\small                        % change to \footnotesize or \scriptsize as needed
\bibliographystyle{abbrvnat}  
\bibliography{references}  
\endgroup

%%%%%%%%%%%%%%%%%%%%%%%%%%%%%%%%%%%%%%%%%%%%%%%%%%%%%%%%%%%%%%%%%

\end{document}